\documentclass[floatfix,twocolumn,showpacs,preprintnumbers,amsmath,amssymb,pra,superscriptaddress,longbibliography]{revtex4-1}
\usepackage{color}
\usepackage[usenames,dvipsnames,svgnames,table]{xcolor}
\usepackage[colorlinks=true,linkcolor=blue,urlcolor=blue,citecolor=blue]{hyperref}
\usepackage{mathtools}
\usepackage{graphicx}
\usepackage{dcolumn}
\usepackage{array}
\usepackage{lipsum}
\usepackage{bm}
\usepackage{subfigure}
\usepackage{amssymb}
\usepackage{multirow}
\usepackage{tabularx}
\usepackage{amsmath}
\usepackage{braket}
\usepackage{csquotes}
\graphicspath{{plots/}}
 \usepackage{lipsum}
\usepackage{mathrsfs}
\usepackage{MnSymbol}

\newcommand{\beq}{\begin{equation}}
\newcommand{\eeq}{\end{equation}}
\newcommand{\bea}{\begin{eqnarray}}
\newcommand{\eea}{\end{eqnarray}}

\usepackage{cleveref}

\begin{document}
\title{Static linear density response from X-ray Thomson scattering measurements:\\ a case study of warm dense beryllium}

\author{Sebastian Schwalbe}
\email{s.schwalbe@hzdr.de}
\affiliation{Center for Advanced Systems Understanding (CASUS), D-02826 G\"orlitz, Germany}
\affiliation{Helmholtz-Zentrum Dresden-Rossendorf (HZDR), D-01328 Dresden, Germany}


\author{Hannah M.~Bellenbaum}

\affiliation{Center for Advanced Systems Understanding (CASUS), D-02826 G\"orlitz, Germany}
\affiliation{Helmholtz-Zentrum Dresden-Rossendorf (HZDR), D-01328 Dresden, Germany}
\affiliation{Institut f\"ur Physik, Universit\"at Rostock, D-18057 Rostock, Germany}


\author{Tilo~D\"oppner}
\affiliation{Lawrence Livermore National Laboratory (LLNL), California 94550 Livermore, USA}

\author{Maximilian~P.~B\"ohme}

\affiliation{Lawrence Livermore National Laboratory (LLNL), California 94550 Livermore, USA}

\author{Thomas~Gawne}

\affiliation{Center for Advanced Systems Understanding (CASUS), D-02826 G\"orlitz, Germany}
\affiliation{Helmholtz-Zentrum Dresden-Rossendorf (HZDR), D-01328 Dresden, Germany}

\author{Dominik~Kraus}
\affiliation{Institut f\"ur Physik, Universit\"at Rostock, D-18057 Rostock, Germany}
\affiliation{Helmholtz-Zentrum Dresden-Rossendorf (HZDR), D-01328 Dresden, Germany}

\author{Michael J.~MacDonald}
\affiliation{Lawrence Livermore National Laboratory (LLNL), California 94550 Livermore, USA}

\author{Zhandos~A.~Moldabekov}

\affiliation{Center for Advanced Systems Understanding (CASUS), D-02826 G\"orlitz, Germany}
\affiliation{Helmholtz-Zentrum Dresden-Rossendorf (HZDR), D-01328 Dresden, Germany}

\author{Panagiotis Tolias}
\affiliation{Space and Plasma Physics, Royal Institute of Technology (KTH), Stockholm, SE-100 44, Sweden}

\author{Jan Vorberger}
\affiliation{Helmholtz-Zentrum Dresden-Rossendorf (HZDR), D-01328 Dresden, Germany}

\author{Tobias Dornheim}
\email{t.dornheim@hzdr.de}

\affiliation{Center for Advanced Systems Understanding (CASUS), D-02826 G\"orlitz, Germany}
\affiliation{Helmholtz-Zentrum Dresden-Rossendorf (HZDR), D-01328 Dresden, Germany}

\begin{abstract}
Linear response theory is ubiquitous throughout physics and plays a central role in the theoretical description of warm dense matter---an extreme state that occurs within compact astrophysical objects and that is traversed on the compression path of a fuel capsule in inertial confinement fusion applications. Here we show how one can relate the static linear density response function to X-ray Thomson scattering (XRTS) measurements, which opens up new possibilities for the diagnostics of extreme states of matter, and for the rigorous assessment and verification of theoretical models and approximations. As a practical example, we consider an XRTS data set of warm dense beryllium taken at the National Ignition Facility [T.~D\"oppner \emph{et al.}, \textit{Nature} \textbf{618}, 270-275 (2023)]. The comparison with state-of-the-art \emph{ab initio} path integral Monte Carlo (PIMC) simulations [T.~Dornheim \emph{et al.}, arXiv:2402.19113] gives us a best estimate of the mass density of $\rho=18\pm6\,$g/cc, which is consistent with previous PIMC and density functional theory based studies, but rules out the original estimate of $\rho=34\pm4\,$g/cc based on a Chihara model fit.
\end{abstract}
\maketitle

\section{Introduction}

Understanding the properties of matter under extreme densities, temperatures and pressures constitutes a highly active frontier at the interface of plasma and solid state physics, quantum chemistry, material science, and a host of other disciplines. In nature, these conditions abound in astrophysical objects such as white~\cite{Kritcher2020,SAUMON20221} and brown dwarfs~\cite{becker} and giant planet interiors~\cite{Benuzzi_Mounaix_2014,guillot2022giant}. In the laboratory, the high-pressure regime is of marked interest for material science, synthesis and discovery~\cite{Kraus2016,Lazicki2021}. A particularly important application is given by the burgeoning field of inertial confinement fusion (ICF)~\cite{Betti2016}, which promises a potential abundance of clean and safe energy in the foreseeable future. In fact, the recent spectacular achievements at the National Ignition Facility (NIF)~\cite{NIF_PRL_2024,Zylstra2022} and at the OMEGA laser facility~\cite{Williams2024} have sparked a surge of research and development activities towards a nuclear fusion power plant~\cite{roadmap}.

An important parameter regime concerns the so-called \emph{warm dense matter} (WDM)~\cite{wdm_book,review}, which can be defined in terms of a few dimensionless parameters that are all of the order of unity~\cite{Ott2018}: i) the density parameter $r_s=d/a_\textnormal{B}$ [where $d$ and $a_\textnormal{B}$ are the Wigner-Seitz and Bohr radius], ii) the degeneracy temperature $\Theta=k_\textnormal{B}T/E_\textnormal{F}$ [where $E_\textnormal{F}$ is the electronic Fermi energy~\cite{quantum_theory}], iii) the coupling parameter $\Gamma=W/K$ [where $W$ and $K$ are the interaction and kinetic energy]. In practice, the fact that $r_s\sim\Theta\sim\Gamma\sim1$ gives rise to an intricate interplay of effects such as partial ionization, partial electronic degeneracy, moderate Coulomb coupling, and strong thermal excitations, which manifest in exotic behavior such as the recently reported free-bound transitions~\cite{boehme2023evidence}. The rigorous theoretical description of WDM thus requires a holistic description, which is notoriously difficult. Arguably, the best contender is given by the direct \emph{ab initio} path integral Monte Carlo (PIMC) technique without any nodal restrictions~\cite{Ceperley1991}, which, however, remains limited to light elements (currently hydrogen~\cite{Dornheim_MRE_2024,Dornheim_JCP_2024} and beryllium~\cite{dornheim2024unraveling,Dornheim_JCP_2024}) and weak to moderate degrees of quantum degeneracy, $\Theta\gtrsim1$~\cite{Bonitz_POP_2024}. A more broadly applicable alternative to PIMC is given by the combination of an \emph{ab initio} density functional theory (DFT) description of the electrons with a semi-classical molecular dynamics propagation of the heavier nuclei within the commonly assumed Born-Oppenheimer approximation. Here, the main challenge is given by the choice of the electronic exchange--correlation (XC) functional~\cite{Moldabekov_JCTC_2024,Bonitz_POP_2024,Karasiev_PRB_2022}, which has to be supplied as a semi-empirical external input.

In addition, the extreme conditions also pose a challenge for the diagnostics of experiments with WDM~\cite{falk_wdm}, as even basic parameters such as the density or temperature cannot be directly measured and have to be inferred indirectly from other observations. The interpretation of experiments thus relies on theoretical models that are generally based on a number of assumptions and approximations~\cite{Dornheim_review}. One of the most important and successful methods for WDM diagnostics is given by x-ray Thomson scattering (XRTS)~\cite{siegfried_review}. The measured intensity can usually be accurately expressed as~\cite{Dornheim_T_follow_up,Gawne_JAP_2024}
\begin{eqnarray}\label{eq:convolution}
I(\mathbf{q},\omega) = S(\mathbf{q},\omega) \circledast R(\omega)\ ,
\end{eqnarray}
i.e., as a convolution between the dynamic structure factor $S(\mathbf{q},\omega)$ that contains important information about the microphysics of the probed sample and the combined source-and-instrument function $R(\omega)$ that follows from the shape of the probing x-ray source and the detector~\cite{Gawne_JAP_2024,gawne2025heartnewxraytracing}. The wave vector $\mathbf{q}$ is determined by the scattering angle, with small and large angles giving access to the collective and non-collective regimes, respectively, while $E=\hbar\omega$ is the energy of the scattered photon. In practice, de-convolving Eq.~(\ref{eq:convolution}) is unstable, which is strongly exacerbated by the experimental noise. Thus, XRTS does not give one direct access to $S(\mathbf{q},\omega)$. Instead, XRTS measurements are usually interpreted by constructing a forward $S(\mathbf{q},\omega)$ model within the Chihara decomposition~\cite{Chihara_1987,Gregori_PRE_2003,boehme2023evidence}, where \emph{a priori} unknown system parameters such as the density are treated as free parameters; these are then inferred by comparing the convolved model intensity with the experimental observation and by finding a best fit~\cite{Kasim_POP_2019}. Clearly, this approach requires the capability to accurately model the dynamic structure factor $S(\mathbf{q},\omega)$, which is difficult.

\begin{figure}\centering
\includegraphics[width=0.49\textwidth]{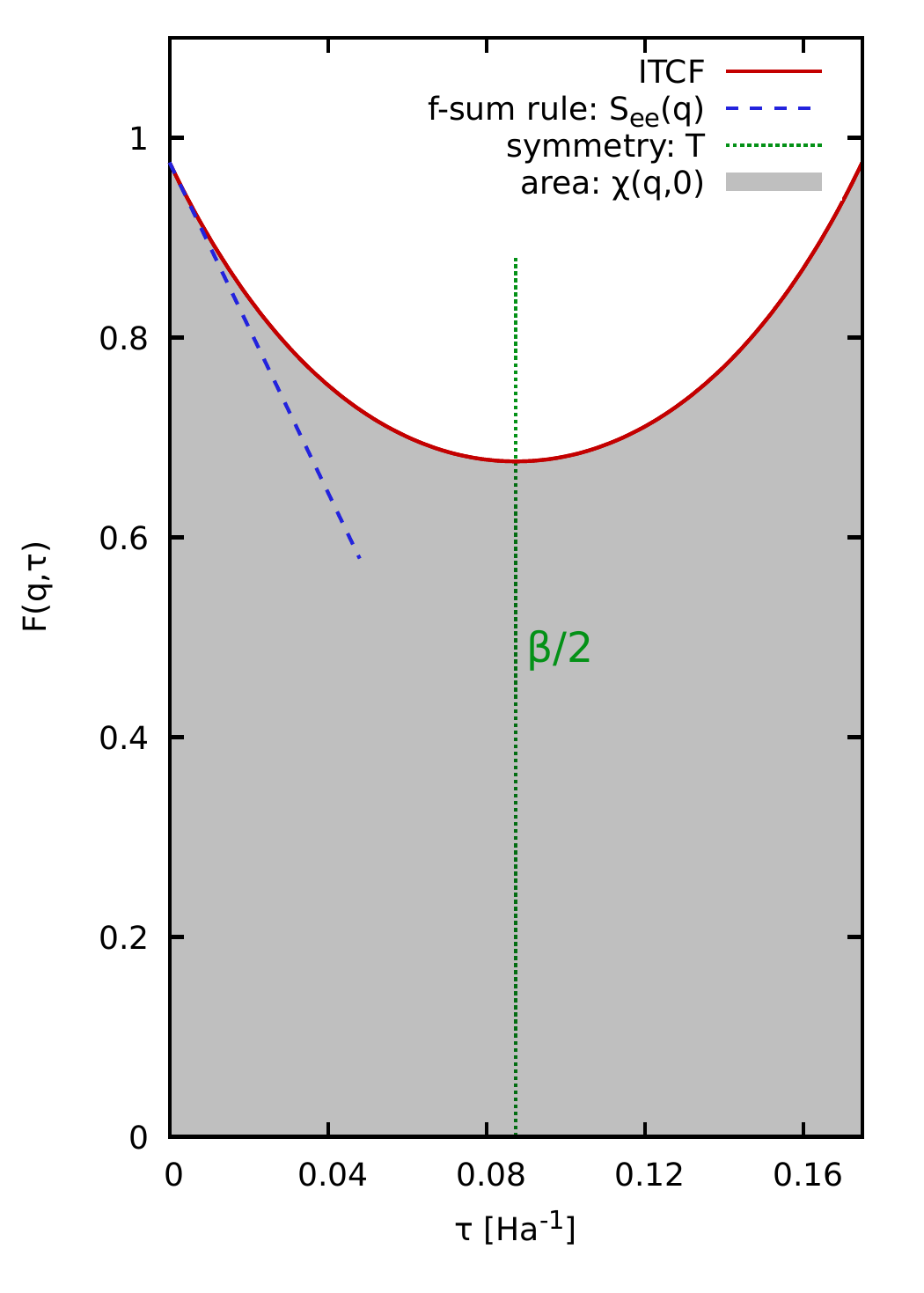}
\caption{\label{fig:ITCF_illustration} The model-free ITCF extraction of various physical properties. Solid red: \emph{ab initio} PIMC results for $F(\mathbf{q},\tau)$ of beryllium at $\rho=7.5\,$g/cc, $T=155.5\,$eV for $q=7.68\,$\AA$^{-1}$~\cite{dornheim2024unraveling}. The symmetry around $\tau=\beta/2$ [dotted green, cf.~Eq.~(\ref{eq:detailed_balance})] determines the temperature; the absolute knowledge of the first derivative of $F(\mathbf{q},\tau)$ around $\tau=0$ from the f-sum rule [dashed blue, cf.~Eq.~(\ref{eq:fsumrule})] gives one access to the normalization and, hence, to $S_{ee}(\mathbf{q})$; the area beneath the ITCF is directly related to the static linear density response $\chi(\mathbf{q},0)$ [shaded black, cf.~Eq.~(\ref{eq:static_chi})]. An overview of practical considerations for the interpretation of measurements is given in Table~\ref{tab:tab}.
}
\end{figure} 

Very recently, it has been suggested to instead consider the two-sided Laplace transform of the dynamic structure factor~\cite{Dornheim_T_2022,Dornheim_MRE_2023,Dornheim_T_follow_up},
\begin{eqnarray}\label{eq:Laplace}
    F(\mathbf{q},\tau) := \mathcal{L}\left[S(\mathbf{q},\omega)\right] = \int_{-\infty}^\infty \textnormal{d}\omega\ S(\mathbf{q},\omega)\ e^{-\tau \hbar \omega}\ ,
\end{eqnarray}
where $F(\mathbf{q},\tau)$ is the imaginary-time density--density correlation function (ITCF). It naturally emerges in Feynman's path integral picture of statistical mechanics~\cite{kleinert2009path} and it contains, by definition, the same information as $S(\mathbf{q},\omega)$, but in a different representation~\cite{Dornheim_MRE_2023,Dornheim_PTR_2023,bellenbaum2025estimatingionizationstatescontinuum}, see Fig.~\ref{fig:ITCF_illustration} for corresponding exact \emph{ab initio} PIMC results for warm dense beryllium. A key advantage of switching to the Laplace domain is the convolution theorem
\begin{eqnarray}\label{eq:convolution_theorem}
    \mathcal{L}\left[S(\mathbf{q},\omega)\right] = \frac{ \mathcal{L}\left[S(\mathbf{q},\omega)\circledast R(\omega)\right]}{ \mathcal{L}\left[R(\omega)\right]}\ ,
\end{eqnarray}
which is remarkably stable with respect to noise in the experimental data, and which gives one direct access to the ITCF. As a first application, it has been demonstrated that the imaginary-time symmetry relation for the ITCF (which is equivalent to the detailed balance relation for the dynamic structure factor)
\begin{equation}\label{eq:detailed_balance}
    F(\mathbf{q},\tau) = F(\mathbf{q},\beta-\tau),  
\end{equation}
where $\beta=1/k_\textnormal{B}T$ is the inverse temperature in energy units, allows the estimation of the temperature of arbitrarily complicated materials in thermal equilibrium without the need for models or approximations~\cite{Schoerner_PRE_2023,dornheim2024unraveling,boehme2023evidence}.
This is illustrated by the dotted green vertical line in Fig.~\ref{fig:ITCF_illustration}, which indicates the minimum in the ITCF around $\tau=\beta/2$. A second application of Eqs.~(\ref{eq:Laplace}) and (\ref{eq:convolution_theorem}) is given by the f-sum rule in the imaginary--time domain~\cite{Dornheim_SciRep_2024},
\begin{eqnarray}\label{eq:fsumrule}
 \left.\frac{\partial}{\partial\tau}F(\mathbf{q},\tau)\right|_{\tau=0} = - \frac{\hbar^2q^2}{2m_e}\ ,
\end{eqnarray}
which determines the slope of the ITCF around $\tau=0$; see the dashed blue line in Fig.~\ref{fig:ITCF_illustration}. In practice, Eq.~(\ref{eq:fsumrule}) allows one to infer the a-priori unknown normalization of the measured XRTS intensity and, in this way, gives one both the electronic static structure factor $S_{ee}(\mathbf{q})=F(\mathbf{q},0)$ and the properly normalized ITCF $F(\mathbf{q},\tau)$. Very recently, Dornheim \emph{et al.}~\cite{dornheim2024modelfreerayleighweightxray} have proposed to combine the experimentally inferred $S_{ee}(\mathbf{q})$ with the ratio of elastic-to-inelastic scattering contributions, which allows the model-free estimation of the Rayleigh weight $W_R(\mathbf{q})=S_{eI}^2(\mathbf{q})/S_{II}(\mathbf{q})$ with $S_{eI}(\mathbf{q})$ and $S_{II}(\mathbf{q})$ the electron--ion and ion--ion static structure factors. In particular, $W_R(\mathbf{q})$ can be readily estimated from DFT-MD or PIMC simulations, and has proven to be very useful for the interpretation of a recent XRTS measurement on warm dense beryllium taken at the NIF~\cite{Tilo_Nature_2023}. Finally, Vorberger \emph{et al.}~\cite{Vorberger_PhysLettA_2024} have recently suggested to use any violation of the detailed balance relation (\ref{eq:detailed_balance})
to quantify possible signals of non-equilibrium, which works particularly well when XRTS measurements are being taken at two distinct scattering angles~\cite{Bellenbaum_APL_2025}.

In the present work, we build upon these ideas and relate the static limit of the dynamic linear density response function $\chi(\mathbf{q},\omega)$ with the area under $F(\mathbf{q},\tau)$, cf.~Eq.~(\ref{eq:static_chi}) below and the shaded black area in Fig.~\ref{fig:ITCF_illustration}. First, this new framework constitutes an alternative, independent way to infer the density (and other unknown parameters) without the need for explicitly dynamic simulations. For example, the static density response $\chi(\mathbf{q},0)$ can be straightforwardly computed from equilibrium DFT-MD simulations~\cite{Moldabekov_JCTC_2023,Dornheim_review}, where the XC functional is the only unknown parameter. Such calculations are substantially more controlled than time-dependent DFT calculations of the full $S(\mathbf{q},\omega)$ or $\chi(\mathbf{q},\omega)$, which require additional and less understood approximations either to the dynamic XC-kernel, or the dynamic XC-potential~\cite{ullrich2011time,marques2012fundamentals}. Second, having XRTS measurements at multiple scattering angles and, therefore, multiple wave vectors $\mathbf{q}$ allows one to assess the accuracy of theoretical models and simulations by checking their consistency with the available data points for a fixed set of parameters. Our approach thus opens up the way for dedicated experimental campaigns that will unambiguously characterize the accuracy of different theoretical models and approximations. 
Third, we note that investigating the electronic density response of WDM is important in its own right~\cite{Dornheim_review,Dornheim_MRE_2024,Bohme_PRL_2022,Bohme_PRE_2023}, and will give new insights into the electronic localization around the ions and related effects such as the heuristic (though often useful) concept of ionization. Finally, the presented approach does not presuppose any kind of decomposition of the measured signal, be it into effectively bound and free electrons or elastic and inelastic contributions to $S_{ee}(\mathbf{q},\omega)$. As a practical example, we re-examine an XRTS measurement on strongly compressed beryllium taken at the NIF~\cite{Tilo_Nature_2023} and find overall consistency in terms of the inferred density both with PIMC calculations of different observables presented in Ref.~\cite{dornheim2024unraveling} and with the recent analysis of the Rayleigh weight presented in Ref.~\cite{dornheim2024modelfreerayleighweightxray}.

The layout of this work is as follows. We shortly discuss linear response theory in the context of XRTS in \cref{subsec:lrt} as well as  the the static linear density response function in \cref{subsec:sldrf}. Practical considerations are given in \cref{sec:pc}. The results of the presented methodology are discussed in \cref{sec:results}. The article concludes with a brief summary and discussion in \cref{sec:summary}.

\section{Theory\label{sec:theory}}

\subsection{Linear response theory \label{subsec:lrt}}

The complete information about the density response of any system to an infinitesimal external perturbation is encoded in the dynamic linear density response function $\chi(\mathbf{q},\omega)$, where $\mathbf{q}$ and $\omega$ are the wave vector and frequency of the external perturbation. From a theoretical perspective, it is convenient to express $\chi(\mathbf{q},\omega)$ as~\cite{Dornheim_review,quantum_theory,kugler1}
\begin{eqnarray}\label{eq:define_G}
    \chi(\mathbf{q},\omega) = \frac{\chi_0(\mathbf{q},\omega)}{1 - \left[v(q) + K_\textnormal{xc}(\mathbf{q},\omega)\right]\chi_0(\mathbf{q},\omega)}\ .
\end{eqnarray}
For a uniform electron gas~\cite{Dornheim_review,Hamann_PRB_2020,dornheim_dynamic}, $\chi_0(\mathbf{q},\omega)$ is given by the temperature-dependent Lindhard function that describes the density response of a noninteracting Fermi gas~\cite{quantum_theory} and the XC-kernel is equivalent to the dynamic local field correction~\cite{kugler1,IIT},
\begin{equation}
K_\textnormal{xc}(\mathbf{q},\omega)=-\frac{4\pi{e}^2}{q^2}G(\mathbf{q,\omega)}\ .
\end{equation}
In this case, $K_\textnormal{xc}(\mathbf{q},\omega)$ contains the full wave-vector and frequency-resolved information about electronic XC-effects and setting $K_\textnormal{xc}(\mathbf{q},\omega)\equiv0$ corresponds to the well-known \emph{random phase approximation} (RPA) that describes the density response on a mean-field level. For a real two-component system, where the electrons and nuclei are treated within the same theoretical framework considering all necessary degrees of freedom, Eq.~(\ref{eq:define_G}) becomes more complicated and involves multiple XC-kernels taking into account correlations between all particle species; see Ref.~\cite{Dornheim_MRE_2024} for a corresponding analysis of warm dense hydrogen based on \emph{ab initio} PIMC simulations. Alternatively, one might stick to the spirit of Eq.~(\ref{eq:define_G}) and replace the reference function $\chi_0(\mathbf{q},\omega)$ by the dynamic Kohn-Sham response function $\chi_\textnormal{KS}(\mathbf{q},\omega)$. In that case, the straightforward interpretation of the XC-kernel breaks down as both $\chi_\textnormal{KS}(\mathbf{q},\omega)$ and $K_\textnormal{xc}(\mathbf{q},\omega)$ contain some information about electronic correlations that is encoded in the respective KS-orbitals~\cite{moldabekov_prr_2023, Moldabekov_JCTC_2023, Moldabekov_jcp_2023}. In practice, little is known about the full dynamic XC-kernel of real materials, and the situation is particularly discouraging at finite temperatures. Moreover, a reasonable XC-kernel must be consistent with $\chi_\textnormal{KS}(\mathbf{q},\omega)$ and, thus, has to depend on the particular choice for the XC-functional; this potentially calls into question the straightforward application of general models that have been constructed e.g.~for a simple uniform electron gas~\cite{dornheim_dynamic,Ruzsinszky_PRB_2020,Kaplan_PRB_2022,Panholzer_PRL_2018}.  

Despite these challenges, linear response theory is ubiquitous throughout WDM theory~\cite{Dornheim_review} and enters calculations of stopping powers~\cite{Cayzac2017,Moldabekov_PRE_2020,Faussurier_POP_2025}, electrical conductivity~\cite{Veysman_PRE_2016}, ionization potential depression~\cite{Zan_PRE_2021}, and a gamut of other applications~\cite{Dornheim_PRB_ESA_2021,zhandos_cpp21,siegfried_review,ernstorfer}. In the context of the present work, a particularly important use case of linear response theory is the interpretation of XRTS experiments via the fluctuation--dissipation theorem~\cite{quantum_theory},
\begin{eqnarray}\label{eq:FDT}
S(\mathbf{q},\omega) = -\frac{\hbar}{\pi n }\frac{\textnormal{Im}\chi(\mathbf{q},\omega)}{1-e^{-\beta\hbar\omega}}\ ,
\end{eqnarray}
which is central to the forward modeling of the XRTS intensity via Eq.~(\ref{eq:convolution}); we repeat here that the main limiting factor with this approach is given by the \emph{a priori} unclear dynamic XC-kernel $K_\textnormal{xc}(\mathbf{q},\omega)$.

To overcome this bottleneck, we propose the utilization of the imaginary-time version of the fluctuation--dissipation theorem~\cite{bowen2,Dornheim_MRE_2023} to determine the static linear density response function,
\begin{eqnarray}\label{eq:static_chi}
    \chi(\mathbf{q},0) = -n \underbrace{\int_0^\beta \textnormal{d}\tau\ F(\mathbf{q},\tau)}_{=:L(\mathbf{q})}\ .
\end{eqnarray}
Here $L(\mathbf{q})$ is the area beneath the ITCF, which can be directly extracted from the measured XRTS signal without the need for model calculations or approximations. In practice, we can compare the experimental value for $L(\mathbf{q})$ to any theoretical method that is capable of computing a static density response function via
\begin{eqnarray}\label{eq:area}
    L(\mathbf{q}) = - \frac{1}{n} \chi(\mathbf{q},0)\ .
\end{eqnarray}
Crucially, Eq.~(\ref{eq:area}) does not require any explicit information about the dynamic XC-kernel. More specifically, the static density response function can be estimated based on results for the single-electron density alone, which makes it easily accessible e.g.~to standard DFT-MD simulations~\cite{Moldabekov_JCTC_2022,Moldabekov_JCTC_2023,Moldabekov_PPNP_2024}, including high-temperature versions such as orbital-free DFT~\cite{Sjostrom_PRB_2013,wesolowski2013recent,Moldabekov_PRB_2023, Moldabekov_Elec_Struc_2025}, extended Kohn-Sham DFT~\cite{Zhang_POP_2016,Blanchet_POP_2020,BLANCHET2022108215}, and spectral quadrature DFT~\cite{SURYANARAYANA2018288,Bethkenhagen_PRE_2023}.

\subsection{Static linear density response function \label{subsec:sldrf}}

The static linear density response function is directly related to the dynamic structure factor via the inverse moment sum rule~\cite{Vitali_PRB_2010,dornheim_dynamic,tkachenko_book}, 
\begin{eqnarray}\label{eq:inverse_moment}
    \chi(\mathbf{q},0) = - \frac{2n}{\hbar}\int_{0}^\infty \textnormal{d}\omega\ \frac{S(\mathbf{q},\omega)}{\omega}\left\{
    1-e^{-\beta\hbar\omega}
    \right\}\ .
\end{eqnarray}
In practice, the evaluation of Eq.~(\ref{eq:inverse_moment}) based on experimental data is precluded by the convolution with $R(\omega)$, cf.~Eq.~(\ref{eq:convolution}). However, Eq.~(\ref{eq:inverse_moment}) is useful a) to test dynamic though approximate data for $S(\mathbf{q},\omega)$ against more reliable data for the static density response $\chi(\mathbf{q},0)$, b) to get additional insights into the physical meaning of $\chi(\mathbf{q},0)$.

In particular, the dynamic structure factor is often expressed as~\cite{Vorberger_PRE_2015}
\begin{eqnarray}\label{eq:define_WR}
    S(\mathbf{q},\omega) = \underbrace{W_R(\mathbf{q}) \delta(\omega)}_{S_\textnormal{el}(\mathbf{q},\omega)} + S_\textnormal{inel}(\mathbf{q},\omega)\ .
\end{eqnarray}
The first term describes the (quasi-)elastic scattering signal with $W_R(\mathbf{q})$ being the Rayleigh weight~\cite{dornheim2024modelfreerayleighweightxray}; it contains contributions both from the atomic form factor and from the free electronic screening cloud, although such a decomposition into effectively free and bound electrons is optional. Similarly, $S_\textnormal{inel}(\mathbf{q},\omega)$ describes all inelastic processes, including free--free, bound--free, and free--bound (see Ref.~\cite{boehme2023evidence}) transitions~\cite{Gregori_PRE_2003}. From Eq.~(\ref{eq:inverse_moment}), it directly follows that the static density response [and, thus, also the area beneath the ITCF, $L(\mathbf{q)}$] are highly sensitive to the elastic feature at $\omega\to0$ and, thus, to the degree of electronic localization around the ions~\cite{Dornheim_MRE_2024}. For completeness, note that a qualitative translation of the individual scattering components to the imaginary-time has been presented very recently by Bellenbaum \emph{et al.}~\cite{bellenbaum2025estimatingionizationstatescontinuum}.

\section{Analyzing XRTS experiments with the ITCF formalism\label{sec:pc}}

\begin{table*}
\caption{\label{tab:tab}Model-free extraction of various observables using the ITCF $F(\mathbf{q},\tau)$ together with the relevant references (for more details on the ITCF, see also Refs.~\cite{Dornheim_MRE_2023,Dornheim_moments_2023,Dornheim_PTR_2023,bellenbaum2025estimatingionizationstatescontinuum}). It is emphasized that observables that require the full ITCF [the half-range $\tau\in[0,\beta/2]$ is sufficient due to the symmetry relation (\ref{eq:detailed_balance})] are generally limited to temperatures above a critical temperature $T_\textnormal{crit}$. The latter is mainly determined by the width (and decay in the wings) of the combined source-and-instrument function $R(\omega)$~\cite{Gawne_JAP_2024}, with higher resolution set-ups~\cite{Gawne_PRB_2024} generally leading to lower $T_\textnormal{crit}$.
}
\begin{ruledtabular}
\begin{tabular}{rcccc}
  Observable  & $\tau$-range & $T$-range & Decomposition & Spectral range\\
    \colrule\\[-1ex]
 Temperature $T$~\cite{Dornheim_T_2022,Dornheim_T_follow_up,Schoerner_PRE_2023,dornheim2024unraveling,Bellenbaum_APL_2025} & $\tau\in[0,\beta/2]$ & $T > T_\textnormal{crit}$ & None & $\omega \gg \omega_R$
 \\[+1ex]
  Non-equilibrium~\cite{Vorberger_PhysLettA_2024,Bellenbaum_APL_2025} & $\tau\in[0,\beta/2]$ & $T > T_\textnormal{crit}$ & None & $\omega \gg \omega_R$
 \\[+1ex]
   Normalization, $S_{ee}(\mathbf{q})$~\cite{Dornheim_SciRep_2024,dornheim2024unraveling} & $\tau\to0$ & all $T$ & None & full
 \\[+1ex]
 Rayleigh weight $W_R(\mathbf{q})$~\cite{dornheim2024modelfreerayleighweightxray} & $\tau\to0$ & all $T$ & $S_\textnormal{el}(\mathbf{q},\omega)$ and $S_\textnormal{inel}(\mathbf{q},\omega)$ & full
 \\[+1ex]
Linear response $\chi(\mathbf{q},0)$, $L(\mathbf{q})$ [this work]  & $\tau\in[0,\beta/2]$ & $T>T_\textnormal{crit}$ & None & full
\end{tabular}
\end{ruledtabular}
\end{table*}

In experiments, the XRTS spectrum is determined for a specific $\omega$ range, which in practice never approaches ($-\infty$,$\infty$). Thus, for practical considerations the two-sided Laplace transform of the dynamic structure
factor can be written with explicit upper and lower integration limits 
\begin{eqnarray}\label{eq:practical_laplace}
    F_{x,y}(\mathbf{q},\tau) := \mathcal{L}_{x,y}\left[S(\mathbf{q},\omega)\right] = \int_{x}^y \textnormal{d}\omega\ S(\mathbf{q},\omega) e^{-\hbar\omega\tau},
    \label{eq:fxy}
\end{eqnarray}
with $x$, $y$ referring to the lower and upper bounds, respectively. We also define the truncated deconvolved ITCF
\begin{eqnarray}\label{eq:practical_deconvolved}
    F^I_{x,y}(\mathbf{q},\tau) := \frac{\mathcal{L}_{x,y}\left[I(\mathbf{q},\omega)\right]}{\mathcal{L}\left[R(\omega)\right]}\ ;
\end{eqnarray}
we assume that $R(\omega)$ is available over a sufficiently large frequency-range so that truncation errors in its Laplace transform are negligible. Obviously, Eq.~(\ref{eq:practical_deconvolved}) converges to the familiar convolution theorem Eq.~(\ref{eq:convolution_theorem}) for $x\to-\infty$, $y\to\infty$. Note that it does not automatically hold that $F^I_{x,y}(\mathbf{q},\tau)=F_{x,y}(\mathbf{q},\tau)$ for any $x,y$, since the convolution theorem requires an infinite integration range. In practice, it holds $F^I_{x,y}(\mathbf{q},\tau)\approx F_{x,y}(\mathbf{q},\tau)$ when $y-x\gg \sigma_R$, i.e., when the integration range is significantly larger than the characteristic width $\sigma_R$ of $R(\omega)$~\cite{Dornheim_T_follow_up}. In other words, having a high resolution set-up at an XFEL facility~\cite{Gawne_PRB_2024,Wollenweber_RSI_2021,Descamps2020,Gawne_Electronic_Structure_2025} reduces the spectral range that is required to extract e.g.~the temperature.

In the following, we discuss how Eqs.~\eqref{eq:practical_laplace} and (\ref{eq:practical_deconvolved}) can be used to determine the temperature and degree of non-equilibrium in \cref{subsec:temperature}, to calculate the normalization constant for the ITCF and the Rayleigh weight in \cref{subsec:normalization}, and to determine the area beneath the ITCF and the static linear density response in \cref{subsec:area}. While some of these aspects have already been reported in earlier works for the NIF beryllium data set~\cite{dornheim2024unraveling,dornheim2024modelfreerayleighweightxray}, we feel that it is worth repeating these here in full to comprehensively discuss and juxtapose different considerations for various ITCF-based observables; see Table~\ref{tab:tab} for a tabulated overview and Table~\ref{tab:partical_considerations_ITCF} for some remarks on integration boundaries and extrapolations.

The considered XRTS data set has been obtained at the NIF for strongly compressed beryllium with a beam energy of $\hbar\omega_0=9\,$keV in a backscattering geometry with a scattering angle of $\theta=120^\circ$; it was first published in the work by D\"oppner \emph{et al.}~\cite{Tilo_Nature_2023}.

\begin{figure}\centering
\includegraphics[width=0.449\textwidth]{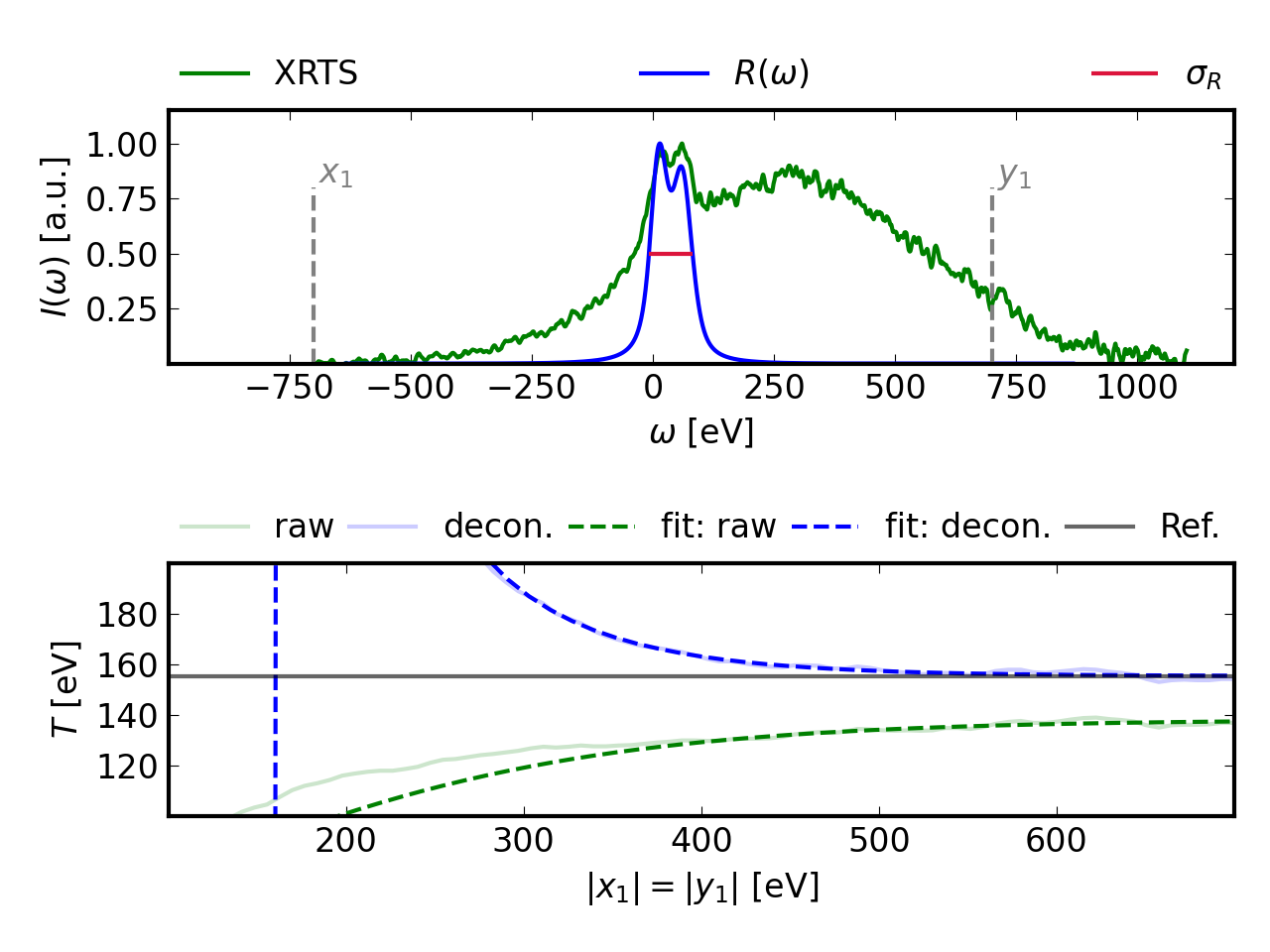}
\caption{\label{fig:beta_half_convergence} The model-free extraction of the temperature: (top) XRTS intensity of strongly compressed beryllium taken at the NIF~\cite{Tilo_Nature_2023} as a function of the photon energy gain;
(bottom) convergence of the ITCF minimum $\tau_\textnormal{min}$, represented as the respective temperature, with the symmetric integration range $x_1=-y_1$. The dashed and shaded curves show the actual temperature estimates and an empirical exponential fit of the same form as Eq.~(\ref{eq:normalization_fit}). Adapted from Ref.~\cite{dornheim2024unraveling}. 
}
\end{figure}

\subsection{Temperature \label{subsec:temperature}}

It is easy to see that $F_{x,y}(\mathbf{q},\tau)$ obeys the symmetry relation Eq.~(\ref{eq:detailed_balance}) for any $x=-y$. To determine the temperature~\cite{Dornheim_T_2022,Dornheim_T_follow_up} or the degree of non-equilibrium~\cite{Vorberger_PhysLettA_2024,Bellenbaum_APL_2025}, we thus choose symmetric integration boundaries, i.e., $x_{1}=-y_{1}$ [see Fig.~\ref{fig:beta_half_convergence} (top)], and compute $F^I_{-x_1,x_1}(\mathbf{q},\tau)$. For the ITCF thermometry approach, we localize the minimum of $F^I_{-x_1,x_1}(\mathbf{q},\tau)$, $\tau_\textnormal{min}(x_1)$, which converges towards $\tau_\textnormal{min}=\beta/2$ upon increasing $x_1$, i.e., when $F_{-x_1,x_1}^I(\mathbf{q},\tau)\approx F_{-x_1,x_1}(\mathbf{q},\tau)$.

The corresponding convergence of $\beta/2$ (converted to temperatures) with $x_1$ is shown in Fig.~\ref{fig:beta_half_convergence} (bottom), where the green and blue curves show results for the unconvolved Laplace transform of the full intensity (i.e., setting $\mathcal{L}\left[R(\omega)\right]\equiv1$) and the properly deconvolved truncated Laplace transform $F^I_{-x_1,x_1}(\mathbf{q},\tau)$, Eq.~(\ref{eq:practical_deconvolved}). Evidently, both curves converge for $x_1\gtrsim 500\,$eV, and the deconvolution using Eq. (\ref{eq:convolution_theorem}) changes the extracted temperature by more than $10\%$ for this data set. Note that an equivalent analysis for this data set has first been presented in Ref.~\cite{dornheim2024unraveling} resulting in a model-free temperature estimate of $T=155.5\pm15\,$eV; the nominal uncertainty here is mainly a consequence of corresponding uncertainties in $R(\omega)$, see also the discussion in Ref.~\cite{boehme2023evidence}. A final consideration concerns the practical upper limit of $x_1$, which we choose from $I(\mathbf{q},\omega)\leq0$, i.e., when the measured intensity signal drops below the noise level. This is an important point, as spurious contributions or noise in the intensity are exponentially enhanced at very negative frequencies due to the $e^{-\tau\hbar\omega}$ factor in the two-sided Laplace transform, cf.~Eq.~(\ref{eq:Laplace}). The appropriate limit for the present example is indicated by the vertical dashed gray lines in Fig.~\ref{fig:beta_half_convergence} (top).

Let us next touch upon the limitations of the thermometry approach. As mentioned earlier, the symmetry relation Eq.~(\ref{eq:detailed_balance}) that is at the heart of the model-free temperature extraction is equivalent to the detailed balance relation~\cite{quantum_theory} $S(\mathbf{q},-\omega)=e^{-\beta\hbar\omega}S(\mathbf{q},\omega)$. Accurately resolving the temperature from the minimum of $F_{-x_1,x_1}(\mathbf{q},\tau)$ thus requires us to clearly resolve enough signal around $\omega=0$, which is particularly challenging for $\omega<0$ due to the additional exponential damping.
Unfortunately, the convolution of $S(\mathbf{q},\omega)$ with $R(\omega)$ limits our access to $F^I_{-x_1,x_1}(\mathbf{q},\tau)$, which only converges towards $F_{-x_1,x_1}(\mathbf{q},\tau)$ for $x_1\gg\sigma_R$.
Therefore, we have to resolve enough signal at $\omega<0$ for $x_1\gg\sigma_R$ for the convolution theorem to properly converge. This is possible i) when the temperature is high enough to make the exponential damping factor $e^{-\beta\hbar\omega}$ in the detailed balance relation sufficiently weak or ii) by implementing a narrow source-and-instrument function with $\sigma_R\ll\beta^{-1}$. Note that determination of the temperature from the ITCF does not require prior normalization of the latter and, therefore, generally makes it unnecessary to integrate over the full dynamic structure factor of the studied system. The same holds for the model-free quantification of non-equilibrium effects proposed by Vorberger \emph{et al.}~\cite{Vorberger_PhysLettA_2024} in terms of the violation of the symmetry equation~(\ref{eq:detailed_balance}), which works particularly well when multiple scattering angles are being probed simultaneously.
Consequently, both the thermometry and non-equilibrium diagnostics work for arbitrarily complicated materials, including high-$Z_A$ (with $Z_A$ being the atomic charge) elements and composite materials.


\begin{figure}\centering
\includegraphics[width=0.449\textwidth]{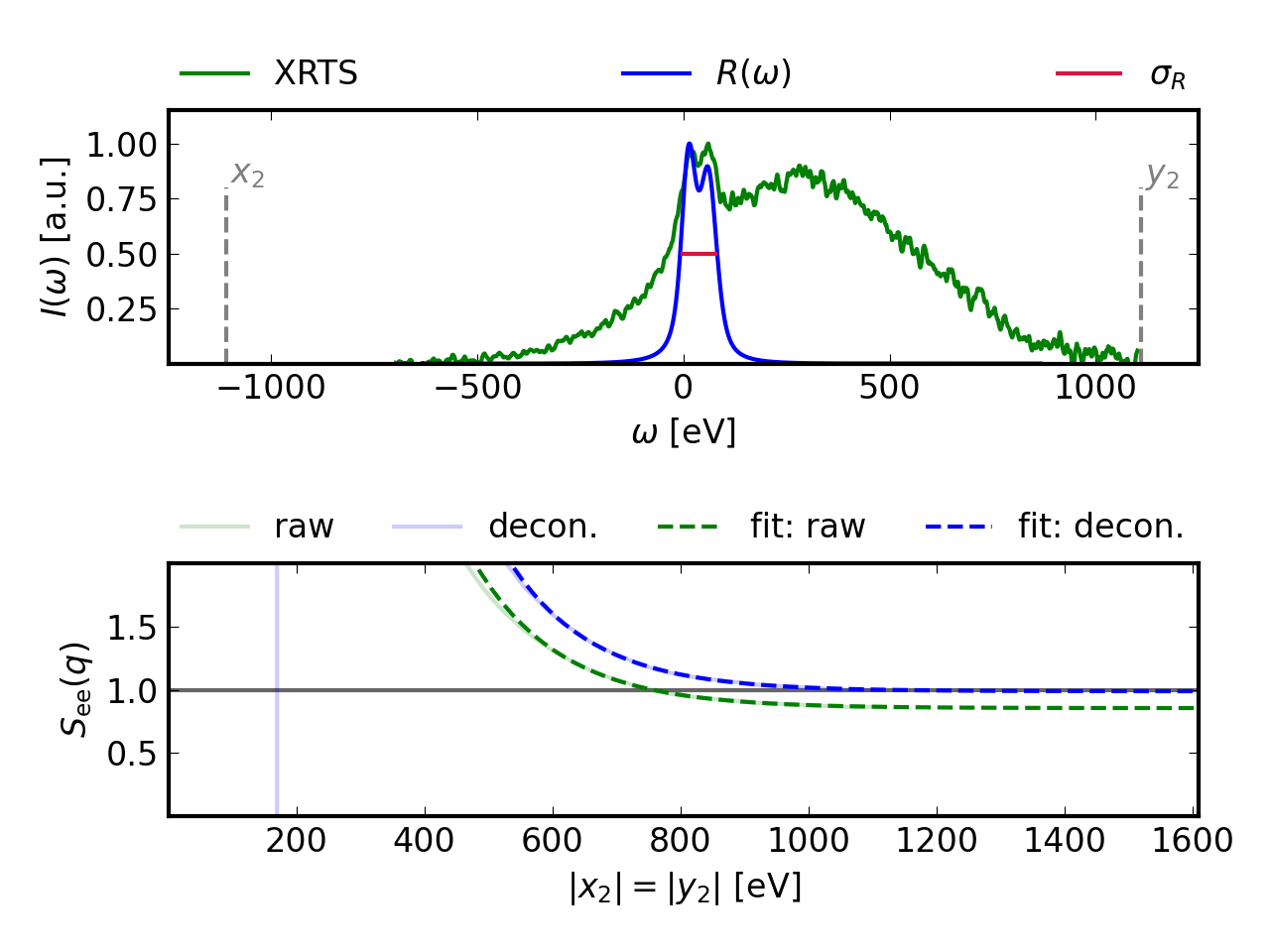}
\caption{\label{fig:norm_convergence} Model-free extraction of the normalization and static structure factor $S_{ee}(\mathbf{q})$: (top) XRTS intensity of strongly compressed beryllium taken at the NIF~\cite{Tilo_Nature_2023} as a function of the photon energy gain; (bottom) convergence of the static structure factor $S_{ee}(\mathbf{q})$ with the integration range $x_2=-y_2$. The dashed and shaded lines show the inferred $S_{ee}(\mathbf{q})$ and an empirical exponential fit, cf.~Eq.~(\ref{eq:normalization_fit}).
}
\end{figure}

\subsection{Normalization\label{subsec:normalization}}

In practice, the measured XRTS intensity [Eq.~(\ref{eq:convolution})] is often only determined up to an a-priori unknown normalization constant. To infer the latter and, in this way, also the proper normalization of $S(\mathbf{q},\omega)$ that is given by the electronic static structure factor $S_{ee}(\mathbf{q}) = F(\mathbf{q},0)$, we require additional information. Dornheim \emph{et al.}~\cite{Dornheim_SciRep_2024} have recently suggested to utilize the imaginary-time version of the well-known f-sum rule~\cite{Dornheim_MRE_2023,Dornheim_PRB_2023}, which determines the first derivative of the ITCF around $\tau=0$, see Eq.~(\ref{eq:fsumrule}). On the one hand, determining the appropriate normalization requires one to integrate over the entire relevant frequency range, including bound-free edges. This becomes increasingly difficult for higher $Z_A$-materials, where the K-edge is getting shifted to larger energies. On the other hand, we only have to resolve the ITCF around $\tau=0$, which means that the exponential enhancement of noise in the up-shifted part of the XRTS spectrum that constitutes the main limitation for the ITCF thermometry approach discussed in the previous section does not play a role here. Indeed, any contributions for large $\omega$ for which  $I(\mathbf{q},\omega)$ might vanish within the given uncertainty range will largely cancel; the integration range can thus be chosen almost arbitrarily large, although it does not make sense to accumulate errors beyond this point. 
Given the robustness of the normalization, we select for convenience again a symmetric integration interval, $x_2=-y_2$.




In Fig.~\ref{fig:norm_convergence}, we demonstrate this procedure for the strongly compressed beryllium NIF data set, similar to the recent analysis of the same spectrum in Ref.~\cite{dornheim2024unraveling}. We observe the onset of convergence for the thus computed electronic static structure factor $S_{ee}(\mathbf{q})$ for $x_2\gtrsim1000\,$eV.
In Ref.~\cite{dornheim2024unraveling}, it has been suggested to perform an empirical exponential fit of the form
\begin{equation}\label{eq:normalization_fit}
    f_{P}(x) = a_{P} + b_{P} e^{c_{P}x}\ , 
\end{equation}
with $a_{P}$, $b_{P}$, $c_{P}<0$ being the fitting parameters and $P$ indicating the targeted property, i.e., $S_{ee}$. 
See, for example, the dashed curves in Fig.~\ref{fig:norm_convergence} (bottom), which fit well to the data; the final result is then given by $S_{ee}(\mathbf{q})=a_{P=S_{ee}}$, i.e., $a_{P=S_{ee},\text{raw}}=0.855$ and  $a_{P=S_{ee},\text{deconv.}}=0.988$ for the raw and deconvolved data, respectively. The proposed fitting form can also be applied to fit the convergence of temperature $T$ [see Fig.~\ref{fig:beta_half_convergence} (bottom)] and the area beneath the ITCF $L(\mathbf{q})$. Moreover, varying the limits of this exponential fit gives one some empirical information about its inherent uncertainty.

Overall, the determination of $S_{ee}(\mathbf{q})$ is very flexible as it only requires knowledge about $F(\mathbf{q},\tau)$ in the vicinity of $\tau=0$. Satisfying the deconvolution theorem such that $F^I_{-x_2,x_2}(\mathbf{q},\tau)\approx F_{-x_2,x_2}(\mathbf{q},\tau)$ is easy as we have to integrate over the entire frequency range anyway. The only requirement is that the spectrum is much broader than the width of $R(\omega)$, which is generally the case except at very small scattering angles. From a physical perspective, it is worth noting that Eq.~(\ref{eq:fsumrule}) is very general and even holds in many non-equilibrium situations, as well as for inhomogeneous samples.

For completeness, we mention that inferring $S_{ee}(\mathbf{q})$ is a key aspect of the model-free extraction of the Rayleigh weight~\cite{dornheim2024modelfreerayleighweightxray},
\begin{eqnarray}\label{eq:Rayleigh_experiment}
    W_R(\mathbf{q}) = \frac{S_{ee}(\mathbf{q})}{1+r^{-1}(\mathbf{q})}\ ,
\end{eqnarray}
where $r(\mathbf{q)}$ is the ratio of elastic-to-inelastic scattering [cf.~Eq.~(\ref{eq:define_WR})]~\cite{Tilo_Nature_2023}:
\begin{eqnarray}\label{eq:ratio}
    r(\mathbf{q}) = \frac{\int_{-\infty}^\infty \textnormal{d}\omega\ S_\textnormal{el}(\mathbf{q},\omega) }{\int_{-\infty}^\infty \textnormal{d}\omega\ S_\textnormal{inel}(\mathbf{q},\omega)}\ .
\end{eqnarray}
We note that $r(\mathbf{q})$ constitutes a standard observable in XRTS experiments and can often be determined with high precision~\cite{dornheim2024unraveling}. At the same time, the theoretical estimates for $r(\mathbf{q})$ require knowledge about correlations between all types of particles in the system, including the electron--electron static structure factor $S_{ee}(\mathbf{q})$; computing the latter is notoriously difficult in particular for DFT-MD simulations, as DFT constitutes an effective single-electron theory.
Eliminating $S_{ee}(\mathbf{q})$ via Eq.~(\ref{eq:Rayleigh_experiment}) based on experimental data removes this obstacle; indeed, computing $W_R(\mathbf{q})$ is a standard practice in DFT-MD, which is highly useful for diagnostic purposes~\cite{dornheim2024modelfreerayleighweightxray}.

\begin{figure}\centering
\includegraphics[width=0.45999\textwidth]{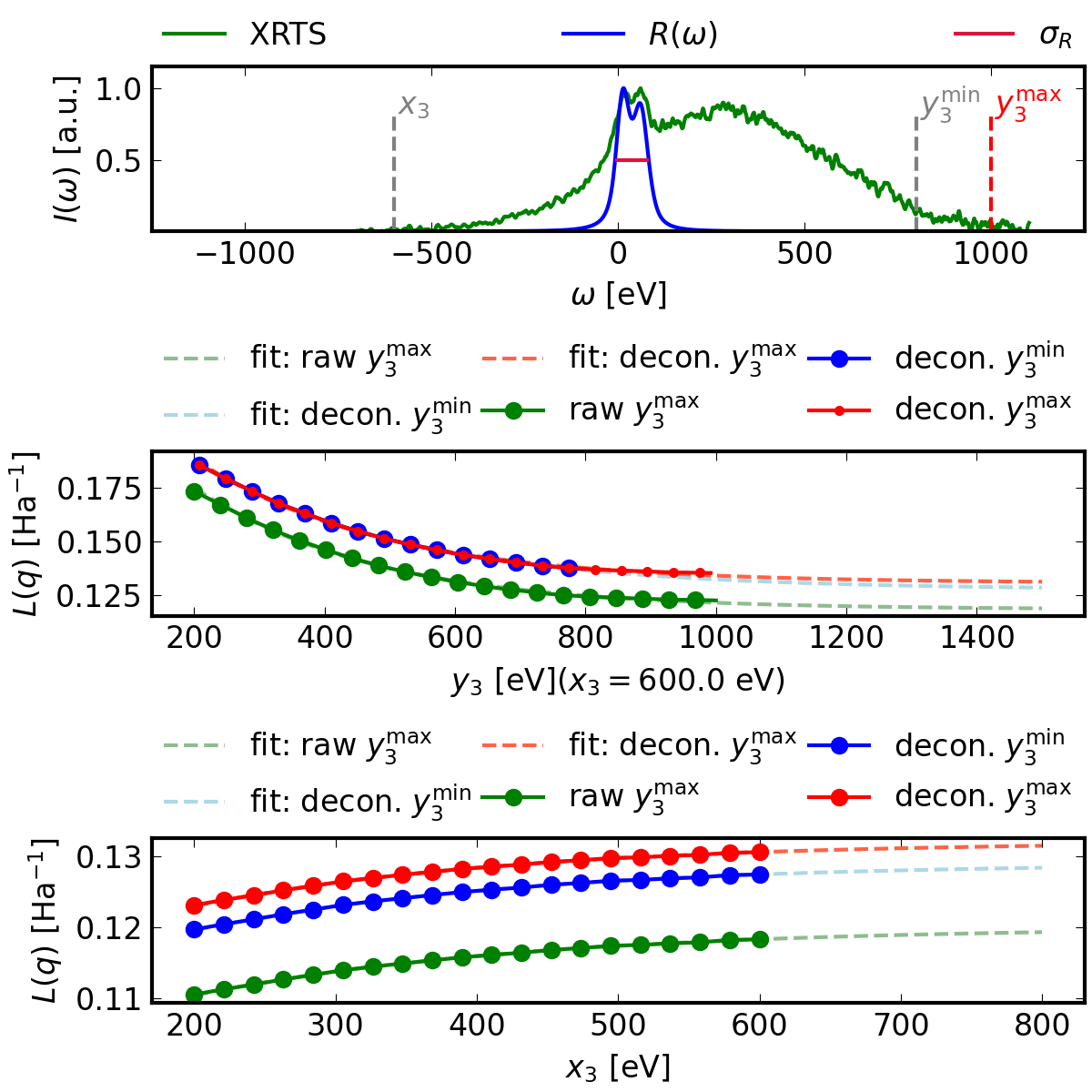}
\caption{\label{fig:area_convergence} Determination of the area beneath the ITCF with a nested two-step convergence procedure:  (top) XRTS intensity of strongly compressed beryllium taken at the NIF~\cite{Tilo_Nature_2023} as a function of the photon energy gain, (middle) convergence of $y_3$ for a fixed value of $x_{3}=-600$ eV, whereas (bottom) shows the global convergence of $|x_3|$, where for a given $|x_3|$, the procedure of (middle) is performed. The respective convergence limit is determined using the empirical fit formula given in Eq.~(\ref{eq:normalization_fit}), see the shaded dashed curves. A similar fit determines the convergence limit of $|x_{3}|$. 
}
\end{figure}

\subsection{Area beneath the ITCF \label{subsec:area}}

Estimating the area beneath the ITCF combines the challenges inherent in the extraction of both previously mentioned observables. First, we have to determine the proper normalization, which requires us to integrate over the entire significant frequency range of the probed material for the given wavenumber $q$.
Second, we have to resolve the entire non-redundant $\tau$-range, i.e., $\tau\in[0,\beta/2]$; this directly follows from the combination of Eq.~(\ref{eq:static_chi}) with the symmetry relation (\ref{eq:detailed_balance}),
\begin{eqnarray}\label{eq:area_beta_half}
    L(\mathbf{q}) = 2\int_0^{\beta/2} \textnormal{d}\tau\ F(\mathbf{q},\tau)\ .
\end{eqnarray}
In practice, Eq.~(\ref{eq:area_beta_half}) is to be preferred to Eq.~(\ref{eq:static_chi}) as the uncertainty in $F^I_{x,y}(\mathbf{q},\tau)$ rapidly increases with $\tau$, see the extensive discussion in Ref.~\cite{Dornheim_T_follow_up}.

The combination of these difficulties requires us to introduce a nested two-step convergence procedure.
First, we select a fixed lower integration limit $x_3$ to exclude the possibility of $I(\mathbf{q},\omega<0)<0$ within the experimental noise; this is a mandatory requirement to prevent the exponential enhancement of noise in the up-shifted part of the XRTS spectrum due to the $e^{-\tau\hbar\omega}$ factor in the two-sided Laplace transform, see also the discussion of the ITCF thermometry approach in Sec.~\ref{subsec:temperature} above:
\begin{eqnarray}
    F_{x_{3},y_{3}}(\mathbf{q},\tau) = \int_{x_{3}}^{y_{3}} d\omega S(\mathbf{q},\omega) e^{-\hbar\omega\tau} .
\end{eqnarray}
Second, we systematically increase the upper integration limit $y_3$ until we reach the limit of the available spectral range. Generally, this is the case when $I(\mathbf{q},\omega>0)$ vanishes within the noise level; for the present spectrum, there appears an unexpected bump at around $\omega\approx850\,$eV, which appears spurious. To factor in the corresponding uncertainty, we first utilize Eq.~(\ref{eq:normalization_fit}) to extrapolate to the limit of $y_3\to\infty$ based on results for $y_3\in[200,800]\,$eV and subsequently perform a second fit for $y_3\in[200,1000]\,$eV; see the blue and red data sets in Fig.~\ref{fig:area_convergence}. While we believe the first fit to be more accurate, we take into account the difference between the two results as an error bar. The integration limits are schematically illustrated by the two vertical gray dashed lines in Fig.~\ref{fig:area_convergence} (top).

To systematically assess the $x_3$ dependence, we repeat the $y_3$ extrapolation for various $x_3$, see Fig.~{\ref{fig:area_convergence}} (bottom). Thus, for a given value of $x_3$ we determine the area as described in the previous paragraph. Those results can be fitted using Eq.~(\ref{eq:normalization_fit}), i.e., $L(\mathbf{q})=a_{P=L_{x_3}}$. 
Using this double convergence procedure regarding $y_3$ and $x_3$, the calculated areas for beryllium NIF data are
$0.129$\,$\text{Ha}^{-1}$ and $0.132$ $\text{Ha}^{-1}$ for the deconvolved spectra assuming maximum values in $y_3$ of $800\,$eV and $1000\,$eV, respectively. We give our final estimate for the area beneath the ITCF of the NIF beryllium data set as $L(\mathbf{q})=0.1305\pm0.005$\,$\text{Ha}^{-1}$, where we have also factored in the uncertainty from the determination of the normalization discussed in the previous section.

Let us finish with a few practical remarks. The integration limits follow the relation $|x_{3}|\leq y_{3}$. 
The lower integration boundary $x_{3}$ can be accessed from the XRTS spectra, i.e., $I(q,x_{3}) < 0 $.
Similar to the determination of the normalization, for the upper boundary, one needs to inspect the right side of the XRTS spectrum for nonphysical signatures and adjust $y_{3}$, see Fig.~{\ref{fig:area_convergence}} (top) respectively. 
The total area beneath the ITCF considering the integration boundaries is then written as 
\begin{equation}
L_{x_{3}y_{3}}(\mathbf{q},0) =  2 \int_{0}^{\beta/2} d\tau \biggl[ \int_{x_{3}}^{y_{3}} d\omega S(\mathbf{q},\omega) e^{-\hbar\omega\tau}\biggr]\ .
\end{equation}

All practical considerations for the model-free extraction of the temperature, the normalization, and the area beneath the ITCF are summarized in Table~\ref{tab:partical_considerations_ITCF}.



\begin{table*}
\caption{Practical considerations for the model-free extraction of various observables using the ITCF $F(\mathbf{q},\tau)$, i.e., integration boundaries for the two-sided Laplace transform and proposed fit functions. }
\begin{ruledtabular}
    \centering
    \begin{tabular}{lcl}
         Observable &  Boundaries & Fit Eq.~(\ref{eq:normalization_fit})\\ \colrule\\[-1ex]
         Temperature & $x_{1} =-y_{1}$ &  $\tau_{\text{min}}=a_{\tau_{\text{min}}}$ \\ \\[-1ex] 
         Normalization & $x_{2}=-y_{2}$ & $S_{ee}(\mathbf{q})=a_{S_{ee}}$ \\ \\[-1ex] 
         Area (this work) & $[x_{3},y_{3}]$ & $L(\mathbf{q})=a_{L,x_{3}}(a_{L,y_{3}})$ \\ \\[-1ex] 
    \end{tabular}
    \label{tab:partical_considerations_ITCF}
\end{ruledtabular}
\end{table*}

\section{Comparison to simulations and models\label{sec:results}}

\subsection{Total area beneath the ITCF\label{subsec:total_area_under_the_ITCF}}

\begin{figure}\centering
\includegraphics[width=0.49\textwidth]{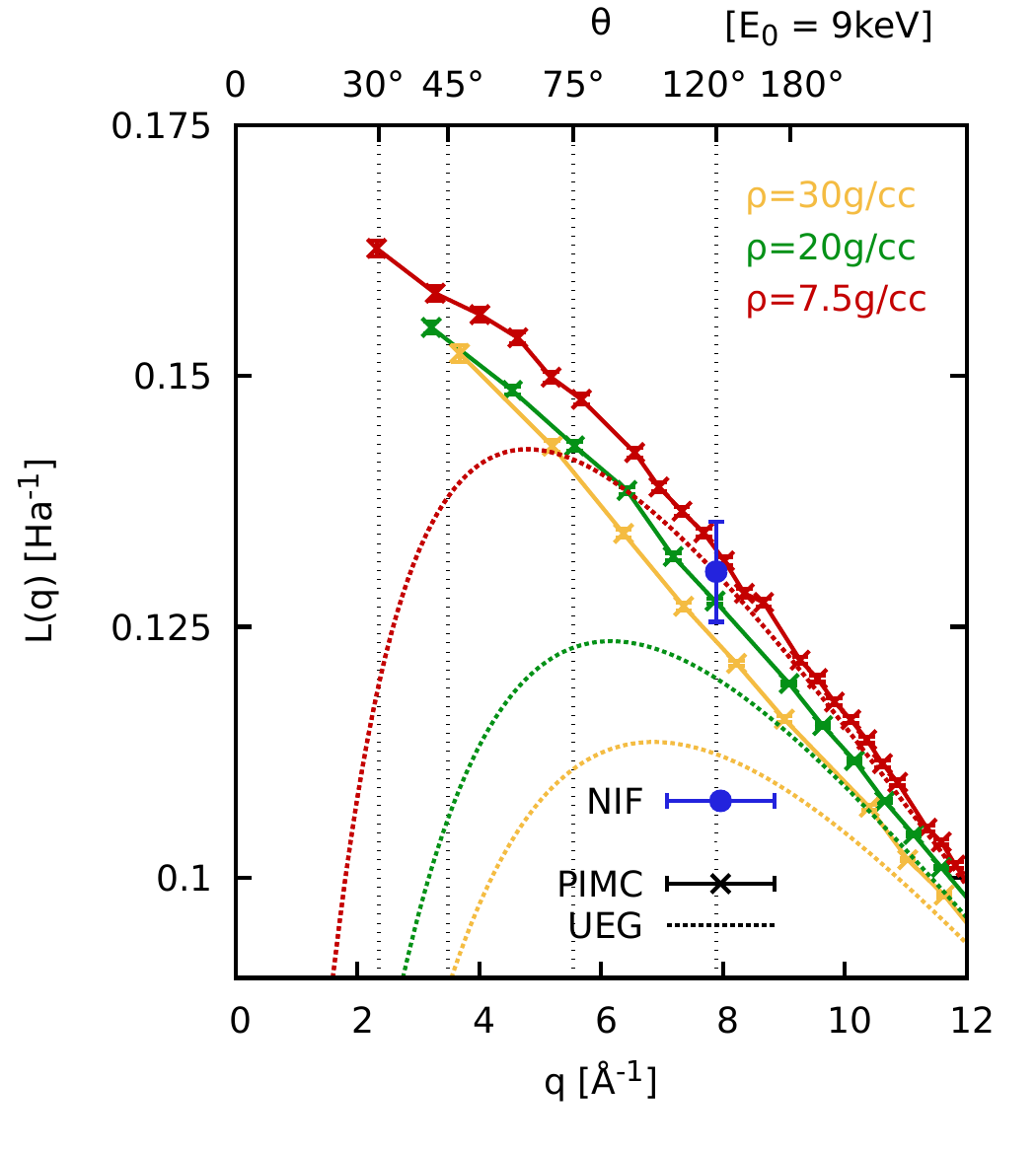}
\caption{\label{fig:Be_area} The area beneath the ITCF $L(\mathbf{q})$ [Eq.~(\ref{eq:area})] for $T=155.5\,$eV~\cite{dornheim2024unraveling} and three different densities. The top and bottom $x$-axis show the scattering angle $\theta$ [assuming a beam energy of $E_0=9\,$keV] and the wavenumber $q$. Crosses: \emph{ab initio} PIMC results computed from $F(\mathbf{q},\tau)$ via Eq.~(\ref{eq:static_chi});
dotted lines: UEG results at the same conditions~\cite{Dornheim_PRB_ESA_2021}; blue point: extracted NIF data point.
}
\end{figure} 

\begin{figure}\centering
\includegraphics[width=0.49\textwidth]{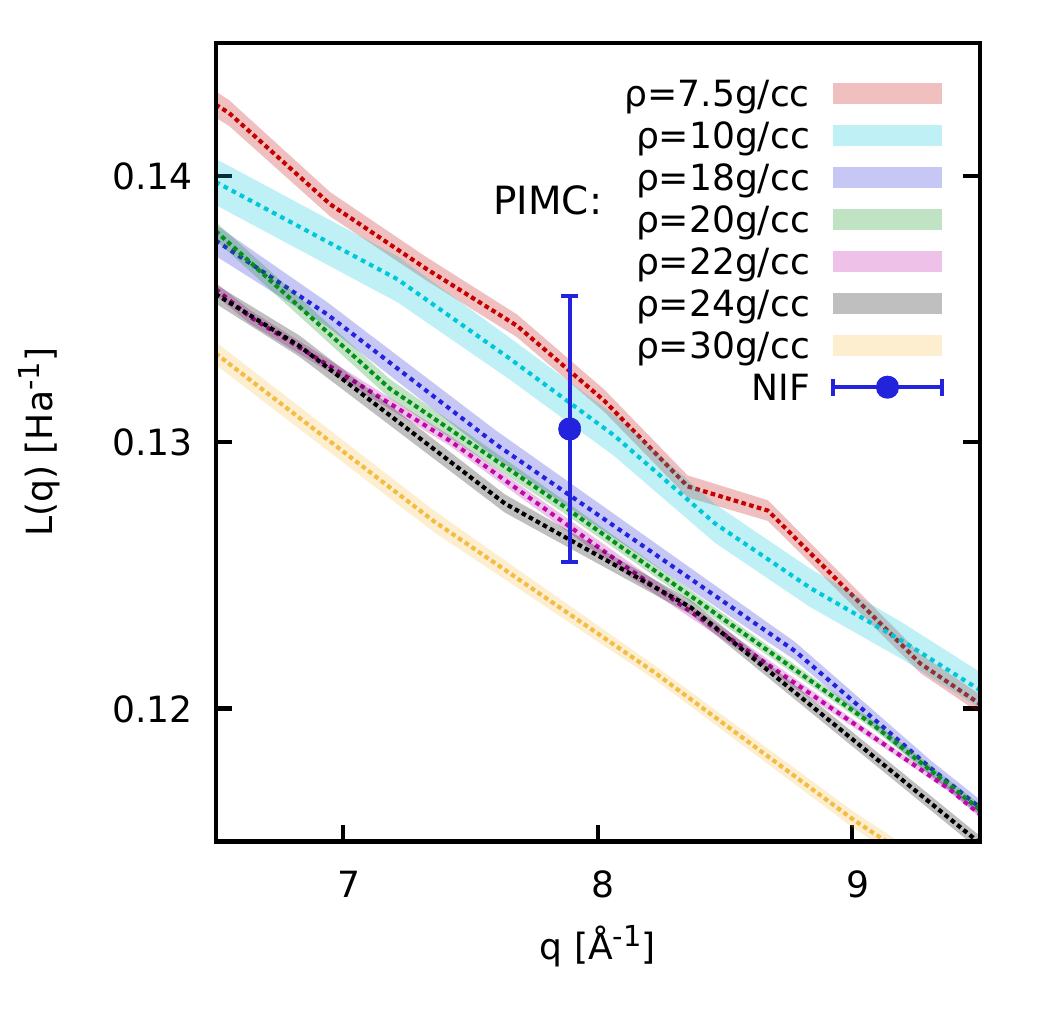}
\caption{\label{fig:Be_area_zoom} The area beneath the ITCF $L(\mathbf{q})$ [Eq.~(\ref{eq:area})] for Be at $T=155.5\,$eV. The dotted curve and shaded areas show \emph{ab initio} PIMC results and their respective uncertainty intervals.
}
\end{figure} 

Having determined both the temperature and the area beneath the ITCF directly from the experimental measurement, we next compare the NIF data point for $L(\mathbf{q})$ to simulation results for different mass densities $\rho$ in Fig.~\ref{fig:Be_area}. The crosses show highly accurate \emph{ab initio} PIMC simulation results that have been computed via Eqs.~(\ref{eq:static_chi}) and (\ref{eq:area}); more details on the PIMC set-up together with some results for the ITCF can be found in Ref.~\cite{dornheim2024unraveling}. In addition, as a reference, we have included results for a uniform electron gas (UEG) at the same conditions computed from the analytical parametrization introduced in Ref.~\cite{Dornheim_PRB_ESA_2021} as the dotted curves. Notably, the UEG curves decay to zero as $q\to0$, which is a direct consequence of perfect screening in the one-component plasma~\cite{kugler_bounds},
\begin{eqnarray}\label{eq:perfect_screening}
    \lim_{q\to0} \chi(\mathbf{q},0) = - \frac{q^2}{4\pi{e}^2}\ .
\end{eqnarray}
In contrast, the elastic feature that is described by the Rayleigh weight $W_R(\mathbf{q})$ [cf.~Eq.~(\ref{eq:define_WR})] increases towards $q\to0$~\cite{dornheim2024modelfreerayleighweightxray}, which explains the differing trend of the PIMC results for beryllium.
For large $q$, on the other hand, the beryllium and UEG results become increasingly similar. Indeed, large wavenumbers mean that one probes the system on increasingly small length scales $\lambda=2\pi/q$, eventually entering the single-particle regime; here, the effects of the nuclei on the density response of the electrons rapidly decrease~\cite{Dornheim_MRE_2024}.

The extracted NIF data point is shown as the blue dot and fits well to PIMC results for $\rho \lesssim25\,$g/cc. This can be seen particularly well in Fig.~\ref{fig:Be_area_zoom}, where we show a magnified segment around the relevant wavenumber.
We note that this is consistent with the previous interpretation of the beryllium XRTS dataset based on the PIMC and DFT-MD based analysis of the Rayleigh weight $W_R(\mathbf{q})$ [$\rho=22\pm2\,$g/cc] presented in Ref.~\cite{dornheim2024modelfreerayleighweightxray}, with the PIMC based analysis of the ratio of elastic-to-inelastic scattering contributions $r(\mathbf{q})$, Eq.~(\ref{eq:ratio}) [$\rho=22\pm2\,$g/cc] presented in Ref.~\cite{dornheim2024unraveling}, and with the average-atom based analysis of $W_R(\mathbf{q})$ [$\rho=20\pm2\,$g/cc] presented in Ref.~\cite{dharmawardana2025xraythomsonscatteringstudies}; the original interpretation of $\rho=34\pm4\,$g/cc, on the other hand, is decisively ruled out.

At the same time, we note that $L(\mathbf{q})$ is not particularly sensitive to different mass densities at these conditions even for relatively small wavenumbers. Indeed, differences between the PIMC datasets for $\rho=7.5\,$g/cc, $\rho=20\,$g/cc and $\rho=30\,$g/cc remain relatively minor even in a forward scattering geometry, see the vertical light grey dotted lines in Fig.~\ref{fig:Be_area} indicating four scattering angles $\theta$ that can be realized at the Gigabar XRTS platform at the NIF.
This is very similar to the insensitivity of the electron--electron static structure factor $S_{ee}(\mathbf{q})$ with respect to $\rho$, which has been reported in Ref.~\cite{dornheim2024unraveling}.
In fact, $L(\mathbf{q})$ and $S_{ee}(\mathbf{q})$ are directly related via the well-known Matsubara series representation~\cite{stls,tolias2024fouriermatsubara}
\begin{eqnarray}\nonumber
    S_{ee}(\mathbf{q}) &=& - \frac{1}{n\beta} \sum_{\ell=-\infty}^\infty \widetilde{\chi}(\mathbf{q},z_\ell)\\\label{eq:series}
    &=& \frac{L(\mathbf{q})}{\beta} - 2 \sum_{\ell=1}^\infty \widetilde{\chi}(\mathbf{q},z_\ell)\ ,
\end{eqnarray}
where $z_\ell=i2\pi\ell/\hbar\beta$ are the discrete imaginary Matsubara frequencies and $\widetilde{\chi}(\mathbf{q},z_\ell)$ the dynamic Matsubara density response function; the second equality directly follows from $\widetilde{\chi}(\mathbf{q},0)=\chi(\mathbf{q},0)$ and the symmetry of the former with respect to the frequency index $\ell$. Recent PIMC based investigations of $\widetilde{\chi}(\mathbf{q},z_\ell)$ of the UEG~\cite{Dornheim_PRB_2024,Dornheim_EPL_2024,Moldabekov_PPNP_2024} have revealed that
the dynamic Matsubara density response is predominantly shaped by single-particle effects for $\ell\neq0$ over the entire $q$-range, which are independent of the density.
In other words, $\chi(\mathbf{q},0)$ [and, hence, $L(\mathbf{q})$]
and $S_{ee}(\mathbf{q})$ only differ by single-particle delocalization effects and, therefore, share their insensitivity to $\rho$.

\begin{figure}\centering
\includegraphics[width=0.49\textwidth]{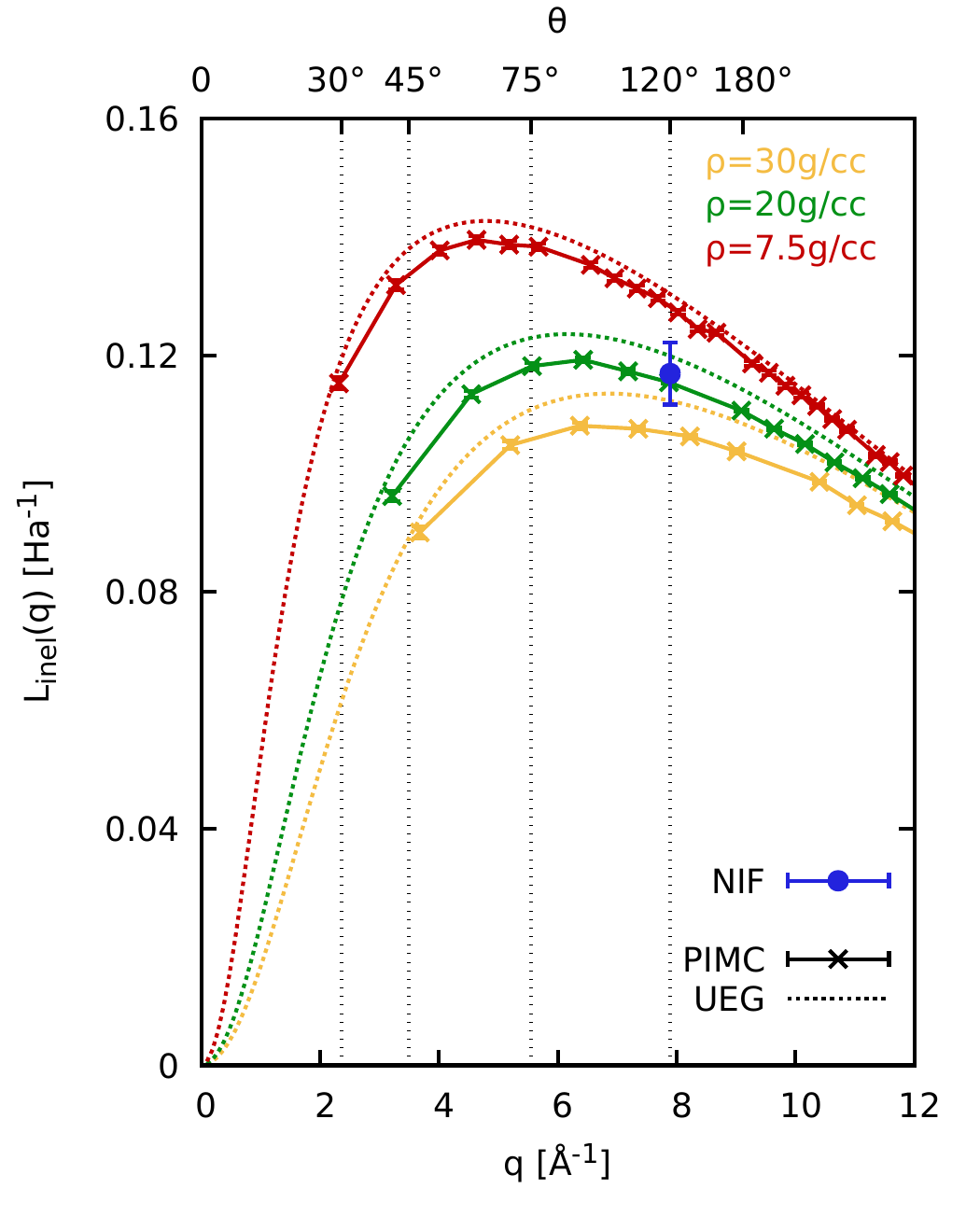}
\caption{\label{fig:inelastic} Inelastic contribution to the area beneath the ITCF $L_\textnormal{inel}(\mathbf{q})$ [Eq.~(\ref{eq:area_inel})], with the same key as in Fig.~\ref{fig:Be_area}. The Rayleigh weight of the experimental data point has been taken from Ref.~\cite{dornheim2024modelfreerayleighweightxray}.
}
\end{figure} 

\begin{figure}\centering
\includegraphics[width=0.49\textwidth]{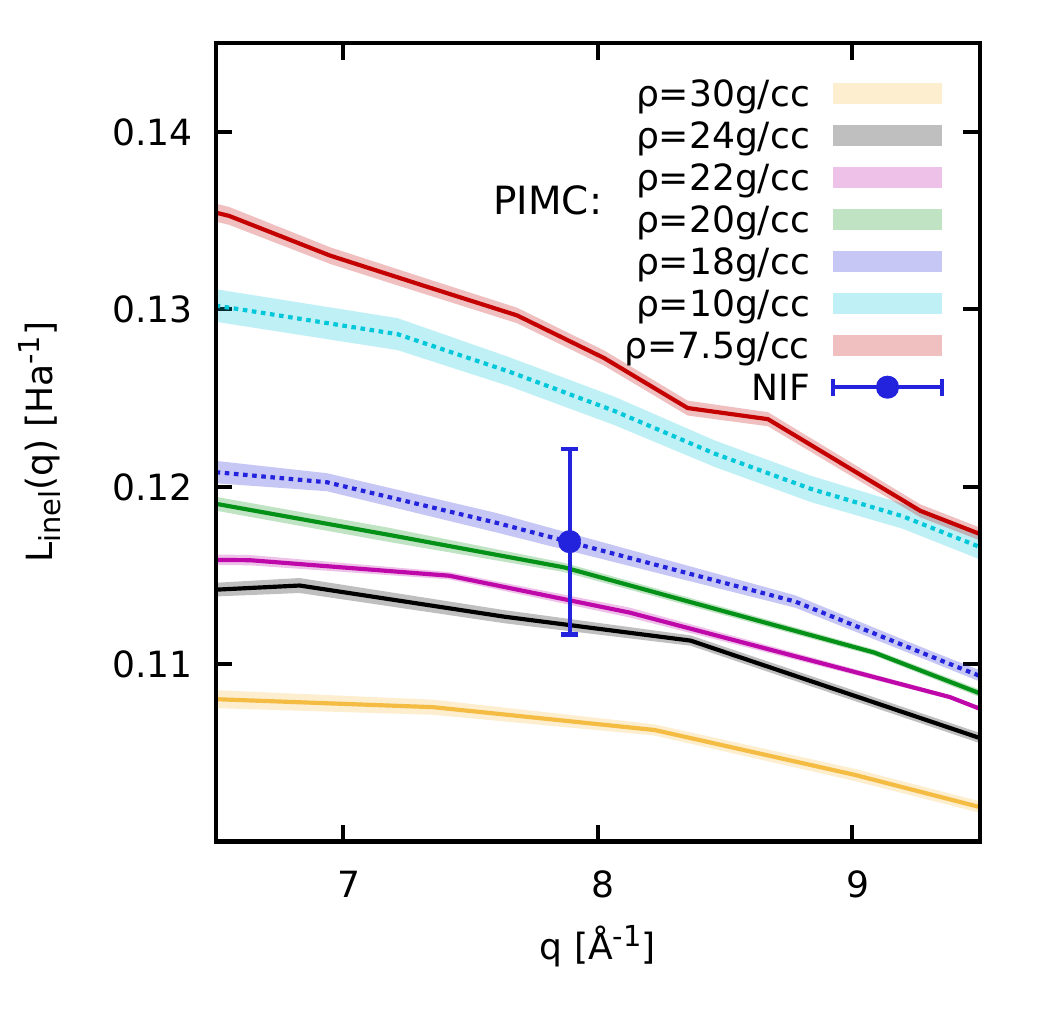}
\caption{\label{fig:shade_zoom_inelastic} Inelastic contribution to the area beneath the ITCF $L_\textnormal{inel}(\mathbf{q})$ [Eq.~(\ref{eq:area_inel})]: magnified segment around the NIF data point in Fig.~\ref{fig:inelastic}.
}
\end{figure}

\subsection{Decomposition into elastic and inelastic contributions\label{subsec:inelastic}}

Aiming to further understand the observed insensitivity of $L(\mathbf{q})$, we propose to decompose it into its elastic and inelastic contributions, 
\begin{eqnarray}\label{eq:area_el}
    L_\textnormal{el}(\mathbf{q}) &=& \beta W_R(\mathbf{q})\ ,\\\label{eq:area_inel}
    L_\textnormal{inel}(\mathbf{q}) &=& 2\int_{0}^{\beta/2}\textnormal{d}\tau\ F_\textnormal{inel}(\mathbf{q},\tau)\\\nonumber
    &=& L(\mathbf{q}) - \beta W_R(\mathbf{q})\ ,
\end{eqnarray}
which follow from consecutively inserting Eq.~(\ref{eq:define_WR}) into the two-sided Laplace transform (\ref{eq:Laplace}) and the imaginary-time version of the fluctuation--dissipation theorem (\ref{eq:static_chi}).
The elastic contribution to the ITCF is simply given by $F_\textnormal{el}(\mathbf{q},\tau)=W_R(\mathbf{q})$, see also the discussion in the recent Refs.~\cite{Dornheim_MRE_2024,bellenbaum2025estimatingionizationstatescontinuum}.
We note that Dornheim \emph{et al.}~\cite{dornheim2024modelfreerayleighweightxray} have used $W_R(\mathbf{q})$ to determine the density of the beryllium XRTS data set and found that it constitutes a good observable for this purpose. Specifically, it was possible to restrict the range of compatible densities to $\pm2\,$g/cc; Dharma-wardana and Klug~\cite{dharmawardana2025xraythomsonscatteringstudies} subsequently found the same sensitivity of $W_R(\mathbf{q})$ in their analysis.
This, by itself, makes the separate analysis of the inelastic contribution also promising.

\begin{figure*}\centering
\includegraphics[width=0.49\textwidth]{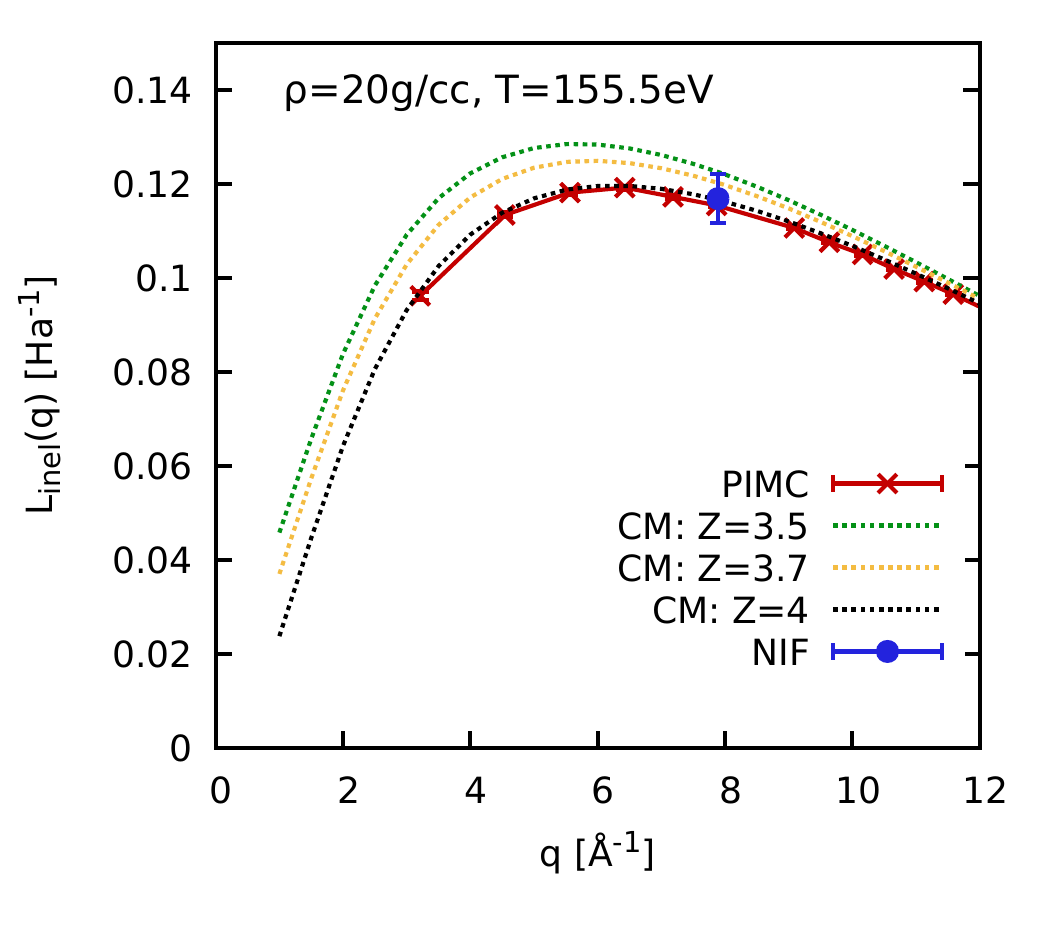}\includegraphics[width=0.49\textwidth]{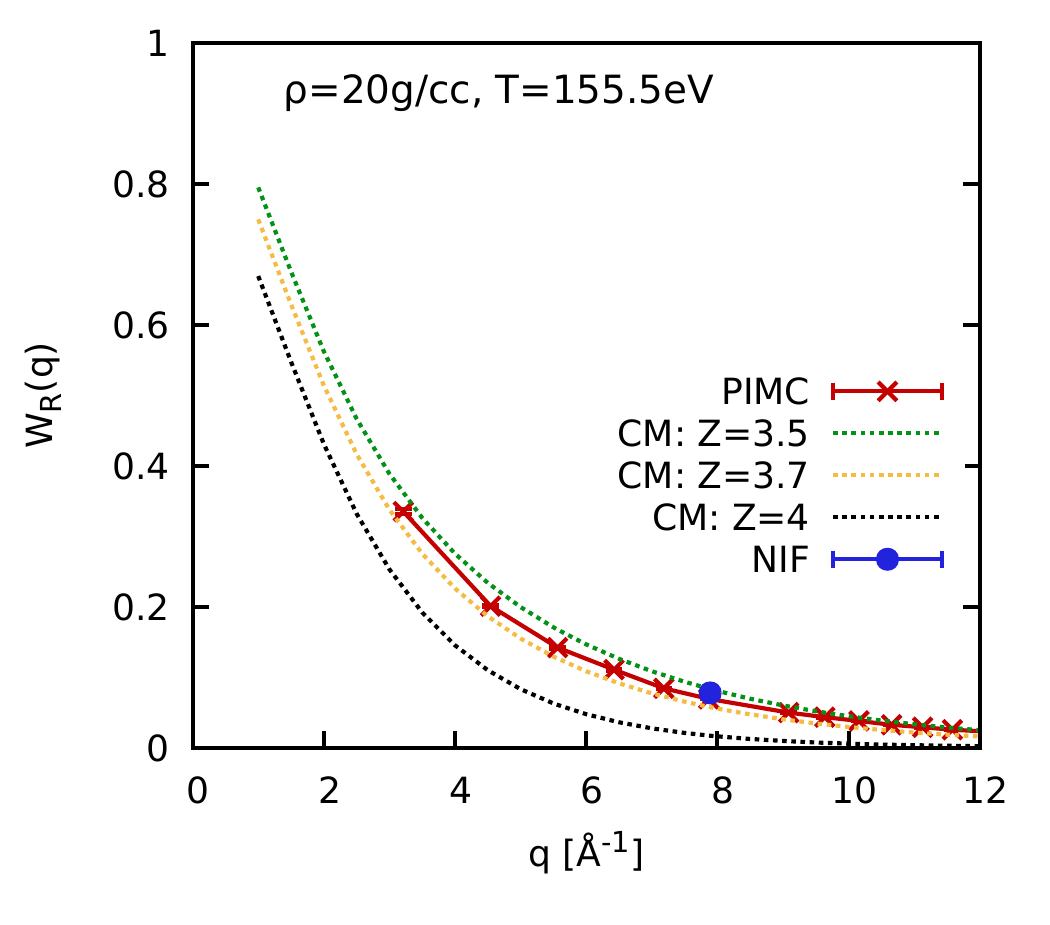}
\caption{\label{fig:I_hate_models} Comparing Chihara models (dotted lines) to \emph{ab initio} PIMC simulations (red crosses) and NIF data (blue dots) for $T=155.5\,$eV and $\rho=20\,$g/cc. The green, yellow and black curves show Chihara results for ionization states of $Z=3.5$, $Z=3.7$, and $Z=4$, respectively. Left: inelastic contribution [Eq.~(\ref{eq:area_inel})]; right: Rayleigh weight $W_R(\mathbf{q})$ [Eq.~(\ref{eq:Rayleigh_experiment})]. The experimental result for $W_R(\mathbf{q})$ has been taken from Ref.~\cite{dornheim2024modelfreerayleighweightxray}.
}
\end{figure*} 

In Fig.~\ref{fig:inelastic}, we show results for $L_\textnormal{inel}(\mathbf{q})$. Evidently, the PIMC results (crosses) are much closer to the UEG curves (dotted lines) than for the full area shown in Fig.~\ref{fig:Be_area} above. This is expected as there is no elastic feature in the dynamic structure factor of the UEG~\cite{quantum_theory,dornheim_dynamic}. The remaining differences between the beryllium and UEG results are a consequence from electronic localization ($Z=3.25-3.6$ have been reported in the literature~\cite{Tilo_Nature_2023,dornheim2024unraveling,dharmawardana2025xraythomsonscatteringstudies}) and transitions between bound and free electronic states~\cite{siegfried_review,boehme2023evidence} that are also absent in the UEG.
Clearly, $L_\textnormal{inel}(\mathbf{q})$ is substantially more sensitive to the density compared to the full area $L(\mathbf{q})$.
The insensitivity of the latter to $\rho$ is thus the result of the near cancellation of two competing trends: for beryllium at the investigated conditions, i) the Rayleigh weight $W_R(\mathbf{q})$ and, hence, $L_\textnormal{el}(\mathbf{q})$ monotonically increase with density (see Fig.~3 in Ref.~\cite{dornheim2024modelfreerayleighweightxray}) whereas ii) $L_\textnormal{inel}(\mathbf{q})$ monotonically decreases with increasing the mass density. 
In practice, we find the largest sensitivity of $L_\textnormal{inel}(\mathbf{q})$ to $\rho$ for intermediate wavenumbers, which would be translated to scattering angles of $\theta\sim45-75^\circ$ for a $9\,$keV probe beam.
Still, we find pronounced differences even at the present backscattering geometry with $\theta=120^\circ$ and $q=7.89$\AA$^{-1}$.

The blue dot corresponds to the NIF XRTS data point, which has been obtained by subtracting the experimental result for $W_R(\mathbf{q})$ that has been reported in Ref.~\cite{dornheim2024modelfreerayleighweightxray} from the present result for the full area $L(\mathbf{q})$, cf.~Eq.~(\ref{eq:area_inel}).
We find very good agreement with the PIMC results for $\rho\sim20\,$g/cc (green crosses), which can be seen particularly well in Fig.~\ref{fig:shade_zoom_inelastic} showing a magnified segment around the experimental region of interest.
The present analysis of the inelastic contribution to the area beneath the ITCF thus gives us a best estimate of the mass density of $\rho=18\pm6\,$g/cc.

\subsection{Comparison with chemical model\label{subsec:chemical}}

Let us conclude our investigation by considering the widely used Chihara model~\cite{Chihara_1987,siegfried_review}.
Owing to their low computational cost, such forward models have been (and are still being) widely used for the interpretation of a host of XRTS experiments~\cite{Gregori_PRE_2003,Fletcher_PRL_2014,Kraus_PRE_2016,kraus_xrts,Tilo_Nature_2023,Poole_PRR_2024}.
In particular, D\"oppner \emph{et al.}~\cite{Tilo_Nature_2023} used a Chihara model to infer the nominal parameters of $T=160\pm20\,$eV and $\rho=34\pm4\,$g/cc for the NIF beryllium XRTS data set that is also being considered in the present work, although transitions of a-priori free electrons to energetically lower bound states were neglected~\cite{boehme2023evidence}. 
We further note that the inherent decomposition into effectively \emph{bound} and \emph{free} populations of electrons automatically introduces an ionization degree~\cite{bellenbaum2025estimatingionizationstatescontinuum}, which, in turn, constitutes an important input for other models and simulations, e.g.~radiation hydrodynamics simulations of inertial fusion energy applications~\cite{drake2018high}.

In the left panel of Fig.~\ref{fig:I_hate_models}, we compare \emph{ab initio} PIMC results for the inelastic area $L_\textnormal{inel}(\mathbf{q})$ at $T=155.5\,$eV and $\rho=20\,$g/cc to Chihara results (dotted curves) with ionization states of $Z=3.5$ (green), $Z=3.7$ (yellow), and $Z=4$ (black).
Here, we used the impulse approximation~\cite{Schumacher_1975} to model the bound-free component of the dynamic structure factor, with the Stewart-Pyatt model~\cite{Stewart_1966} used to estimate ionization potential depression (IPD).
The free-free contribution is estimated by combining the random phase approximation (RPA)~\cite{pines,Bohm_1953} with short distance correlation effects that are accounted for using the local field correction derived in the effective static approximation by Dornheim ~\emph{et al.}~\cite{Dornheim_PRB_ESA_2021,Dornheim_PRL_2020_ESA}.
The static ion-ion structure factors are obtained using an HNC solver~\cite{Wuensch_2008_hnc}.
Screening cloud and form factors are also required to estimate the Rayleigh weight that are modeled using a Debye-H\"uckel~\cite{Wuensch_2008_hnc} pseudo-potential and Pauling-Sherman hydrogenic approximation~\cite{Pauling_1932} respectively. To obtain the area under the (inelastic) ITCF, we have first computed the corresponding DSF $S_\textnormal{inel}(\mathbf{q},\omega)$ over a broad frequency range of the order of $\Delta\omega\sim10\,$keV and subsequently numerically integrated over $\omega$ to obtain $F_\textnormal{inel}(\mathbf{q},\tau)$ and over $\tau$ to obtain $L_\textnormal{inel}(\mathbf{q})$.

First, we observe that the inelastic contribution to the total area beneath the ITCF (and, hence, the predicted magnitude of the static linear density response) systematically drops with increasing $Z$. Interestingly, we only find agreement with the PIMC curve for $Z=4$, which, however, is not realistic at these conditions~\cite{Tilo_Nature_2023,dornheim2024unraveling,dharmawardana2025xraythomsonscatteringstudies}.

In the right panel of Fig.~\ref{fig:I_hate_models}, we show the corresponding Rayleigh weight $W_R(\mathbf{q})$, which determines the elastic contribution, cf.~Eq.~(\ref{eq:area_el}). Here, the Chihara model matches the PIMC curve for $Z\approx3.6$, which is consistent with the PIMC based inference of the ionization state presented in Ref.~\cite{dornheim2024unraveling}.
The Chihara model thus is not capable of consistently matching the different components of the PIMC data (or of the experimental measurement) for a single set of parameters.
We hope that the presented decomposition will give further insights into potential model errors in the individual contributions to the Chihara model, and, in this way, will help to guide the development of improved models.

\section{Summary and discussion \label{sec:summary}}

In this work, we have proposed to consider a new observable for the interpretation of XRTS experiments: the area beneath the deconvolved ITCF $F(\mathbf{q},\tau)$, $L(\mathbf{q})$, which is directly proportional to the static linear density response function $\chi(\mathbf{q},0)$. This idea nicely complements previous ITCF-based aspects of XRTS analysis such as extracting the temperature (or degree on non-equilibrium) from the imaginary-time symmetry relation (\ref{eq:detailed_balance}), obtaining the absolute intensity and electronic static structure factor $S_{ee}(\mathbf{q})$ from the f-sum rule (\ref{eq:fsumrule}) that determines the $\tau-$derivative of the ITCF around $\tau=0$, and estimating the Rayleigh weight by further decomposing the XRTS signal into its elastic and inelastic contributions.

Following a detailed discussion of various practical aspects of the ITCF-based extraction of different properties, we have demonstrated on the example of a recent XRTS dataset of strongly compressed beryllium obtained at the NIF~\cite{Tilo_Nature_2023} that it is indeed possible to infer $L(\mathbf{q})$ with good accuracy. The comparison with state-of-the-art \emph{ab initio} PIMC results from Ref.~\cite{dornheim2024unraveling} has revealed good agreement between the experiment and simulations for $\rho\lesssim25\,$g/cc. This is consistent with previous PIMC and DFD-MD based investigations of other observables~\cite{dornheim2024unraveling,dornheim2024modelfreerayleighweightxray} as well as with the average atom calculations from Ref.~\cite{dharmawardana2025xraythomsonscatteringstudies}, but decisively rules out the nominal value of $\rho=34\pm4\,$g/cc reported in the original paper based on a modified Chihara model~\cite{Tilo_Nature_2023}. At the same time, we conclude that the full area $L(\mathbf{q})$ is substantially less sensitive to $\rho$ than the Rayleigh weight $W_R(\mathbf{q})$ considered in Refs.~\cite{dornheim2024modelfreerayleighweightxray,dharmawardana2025xraythomsonscatteringstudies}.

\begin{table*}
\caption{Overview of various density estimates of the NIF beryllium XRTS data set by D\"oppner \emph{et al.}~\cite{Tilo_Nature_2023}. }
\begin{ruledtabular}
    \centering
    \begin{tabular}{rll}
         Estimate &  Technique & Reference\\ \colrule\\[-1ex]
          $\rho=34\pm4\,$g/cc &  Chihara model fit to $S(\mathbf{q},\omega)$ & D\"oppner \emph{et al.}~\cite{Tilo_Nature_2023} \\ \\[-1ex] 
         $\rho=22\pm2\,$g/cc & PIMC estimation of $r(\mathbf{q})$ [Eq.~(\ref{eq:ratio})] & Dornheim \emph{et al.}~\cite{dornheim2024unraveling} \\ \\[-1ex] 
         $\rho=22\pm2\,$g/cc & DFT-MD and PIMC estimation of $W_R(\mathbf{q})$ [Eq.~(\ref{eq:Rayleigh_experiment})] & Dornheim \emph{et al.}~\cite{dornheim2024modelfreerayleighweightxray} \\ \\[-1ex] 
         $\rho=20\pm2\,$g/cc & average atom estimation of $W_R(\mathbf{q})$ [Eq.~(\ref{eq:Rayleigh_experiment})] & Dharma-wardana and Klug~\cite{dharmawardana2025xraythomsonscatteringstudies} \\ \\[-1ex] 
         $\rho=18\pm6\,$g/cc & PIMC estimation of $L_\textnormal{inel}(\mathbf{q})$ [Eq.~(\ref{eq:area_inel})] & this work\\ \\[-1ex] 
    \end{tabular}
    \label{tab:overview}
\end{ruledtabular}
\end{table*}

To understand the origin of this insensitivity, we have further decomposed $L(\mathbf{q})$ into its elastic (essentially given by the aforementioned Rayleigh weight) and inelastic components. Interestingly, these contributions exhibit opposite trends with respect to $\rho$ that compensate one another to some degree in the full area. Considering the inelastic contribution separately, we have found agreement with our PIMC results for $\rho=18\pm6\,$g/cc, which is fully compatible with previous PIMC based estimates; see Tab~\ref{tab:overview} for a complete overview. This further highlights the great value of \emph{ab initio} PIMC simulations for the interpretation of XRTS experiments and the unprecedented degree of consistency without the need for empirical input parameters such as an ionization degree or exchange--correlation functional. In contrast, widely used Chihara models are not capable of matching both elastic and inelastic components simultaneously with a single set of parameters.

We expect the estimation of the area beneath the ITCF to be valuable for a variety of future projects. First, similar to the Rayleigh weight, $L(\mathbf{q})$ and the directly related density response function $\chi(\mathbf{q},0)$ constitute excellent observables for simulation methods as they do not involve any additional dynamic information such as the dynamic XC-kernel or dynamic XC-potential in different formulations of time-dependent DFT, which have to be approximated in practice. Indeed, Moldabekov \emph{et al.}~\cite{Moldabekov_JCTC_2023,Moldabekov_JCTC_2022,Moldabekov_PPNP_2024} have demonstrated that $\chi(\mathbf{q},0)$ can be computed very accurately from static equilibrium DFT simulations by perturbing the system and subsequently measuring its density response in the limit of small perturbation amplitudes $A$.
An alternative ansatz is given by the powerful density stiffness theorem~\cite{Dornheim_free_energy,Moldabekov_PPNP_2024}
\begin{eqnarray}
    \Delta f(\mathbf{q},A) = \frac{A^2}{n}\chi(\mathbf{q},0)\ ,
\end{eqnarray}
that relates a corresponding shift in the free energy (per particle) $\Delta f$ to $\chi(\mathbf{q},0)$. We note that both the induced density and change in the free energy are particularly suitable observables for DFT, with the XC-functional being the only approximation. Measuring $L(\mathbf{q})$ for multiple scattering angles will thus open up the way to systematically benchmark the accuracy of thermal DFT simulations with different XC-models.
Furthermore, the free energy is also very suitable for novel PIMC simulation approaches~\cite{dornheim2024eta,dornheim2025fermionicfreeenergiestextitab} that deal more efficiently with limitations due to the fermion sign problem~\cite{dornheim_sign_problem}.

A second enticing future application is given by the possibility to first infer $\chi(\mathbf{q},0)$ and subsequently invert Eq.~(\ref{eq:define_G}) to infer the static XC-kernel $K(\mathbf{q},0)$. In this context, we mention the unique possibilities facilitated by the combination of ultrahigh resolution XRTS set-ups~\cite{Wollenweber_RSI_2021,Descamps2020,Gawne_PRB_2024} combined with high repetition rates at modern XFEL facilities such as the European XFEL in Germany~\cite{Tschentscher_2017}.
A particularly promising route is given by isochoric x-ray heating, which means that the density will already be known with very high precision, thus further reducing any uncertainties in $\chi(\mathbf{q},0)$ and, hence, the inferred XC-kernel $K_\textnormal{xc}(\mathbf{q},0)$.

Other works might include the estimation of the higher-order frequency moments of the dynamic structure factor~\cite{Dornheim_moments_2023,Dornheim_MRE_2023}
\begin{eqnarray}\\\nonumber
    \braket{\omega^l} &=& \int_{-\infty}^\infty \textnormal{d}\omega\ S(\mathbf{q},\omega)\ \omega^l \\
    &=& \left(-\frac{1}{\hbar}\right)^l \left. \frac{\partial^l}{\partial\tau^l}F(\mathbf{q},\tau)\right|_{\tau=0} \label{eq:moments}\ ,
\end{eqnarray}
which can be computed from higher-order $\tau$-derivatives of the ITCF at $\tau=0$, for $l\geq1$.
Indeed, Dornheim \emph{et al.}~\cite{Dornheim_moments_2023} have recently demonstrated that it is possible to infer the frequency moments from polynomial fits to the ITCF with high accuracy based on \emph{ab initio} PIMC simulations of the warm dense UEG.
Finally, we mention the possibility to infer the full dynamic Matsubara density response from the ITCF
via~\cite{tolias2024fouriermatsubara,Dornheim_PRB_2024,Dornheim_EPL_2024}
\begin{eqnarray}\label{eq:Fourier_Matsubara}
    \widetilde{\chi}(\mathbf{q},z_\ell) = -2n \int_0^{\beta/2}\textnormal{d}\tau\ F(\mathbf{q},\tau)\ \textnormal{cos}\left(i\hbar{z}_\ell \tau\right)\ ,
\end{eqnarray}
which has given novel insights into the impact of dynamic XC-correlation effects onto the structural properties of the UEG.

\section*{Acknowledgments}
This work was partially supported by the Center for Advanced Systems Understanding (CASUS), financed by Germany’s Federal Ministry of Education and Research (BMBF) and the Saxon state government out of the State budget approved by the Saxon State Parliament. This work has received funding from the European Research Council (ERC) beneath the European Union’s Horizon 2022 research and innovation programme
(Grant agreement No. 101076233, "PREXTREME"). 
Views and opinions expressed are however those of the authors only and do not necessarily reflect those of the European Union or the European Research Council Executive Agency. Neither the European Union nor the granting authority can be held responsible for them. This work has received funding from the European Union's Just Transition Fund (JTF) within the project \emph{R\"ontgenlaser-Optimierung der Laserfusion} (ROLF), contract number 5086999001, co-financed by the Saxon state government out of the State budget approved by the Saxon State Parliament. 
This work of M.P.B., T.D. and M.J.M. was performed beneath the auspices of the U.S. Department of Energy by Lawrence Livermore National Laboratory under Contract No.~DE-AC52-07NA27344 and supported by Laboratory Directed Research and Development (LDRD) Grants No. 24-ERD-044 and 25-ERD-047.

Computations were performed on a Bull Cluster at the Center for Information Services and High-Performance Computing (ZIH) at Technische Universit\"at Dresden, at the Norddeutscher Verbund f\"ur Hoch- und H\"ochstleistungsrechnen (HLRN) under grant mvp00024, and on the HoreKa supercomputer funded by the Ministry of Science, Research and the Arts Baden-W\"urttemberg and
by the Federal Ministry of Education and Research.

\bibliography{bibliography}
\end{document}